**Machine learning accelerates parameter optimization and uncertainty assessment of a land surface model**

**Running title: Uncertainty quantification of LSMs**

Yohei Sawada[1,2,3],

[1] Institute of Engineering Innovation, the University of Tokyo, Tokyo, Japan

[2] Meteorological Research Institute, Japan Meteorological Agency, Tsukuba, Japan

[3] RIKEN Center for Computational Science, Kobe, Japan

Corresponding author: Y. Sawada, Institute of Engineering Innovation, the University of Tokyo, Tokyo, Japan, 2-11-6, Yayoi, Bunkyo-ku, Tokyo, Japan, yohei.sawada@sogo.t.u-tokyo.ac.jp

**Abstract**

The performance of land surface models (LSMs) significantly affects the understating of atmospheric and related processes. Many of the LSMs' soil and vegetation parameters were unknown so that it is crucially important to efficiently optimize them. Here I present a globally applicable and computationally efficient method for parameter optimization and uncertainty assessment of the LSM by combining Markov Chain Monte Carlo (MCMC) with machine learning. First, I performed the long-term (decadal scales) ensemble simulation of the LSM, in which each ensemble member has different parameters' values, and calculated the gap between simulation and observation, or the cost function, for each ensemble member. Second, I developed the statistical machine learning based surrogate model, which is computationally cheap but accurately mimics the relationship between parameters and the cost function, by applying the Gaussian process regression to learn the model simulation. Third, we applied MCMC by repeatedly driving the surrogate model to get the posterior probabilistic distribution of parameters. Using satellite passive microwave brightness temperature observations, both synthetic and real-data experiments in the Sahel region of west Africa were performed to optimize unknown soil and vegetation parameters of the LSM. The primary findings are (1) the proposed method is 50,000 times as fast as the direct application of MCMC to the full LSM; (2) the skill of the LSM to simulate both soil moisture and vegetation dynamics can be improved; (3) I successfully quantify the characteristics of equifinality by obtaining the full non-parametric probabilistic distribution of parameters.

**Key Points**
1. Machine learning realizes a globally applicable method for parameter optimization and uncertainty assessment of the LSMs
2. Non-parametric probabilistic distribution of parameters can be obtained by assimilating satellite observation.

**1. Introduction**

Land surface modeling is fundamental technology to understand, monitor, and predict terrestrial water, energy, and carbon cycles. As a first land surface model (LSM), Manabe (1969) developed the simple bucket model in which heterogeneity of soil, vegetation, and topography was neglected. For many years, LSMs have been evolving from the simple bucket model by including physiological processes (e.g., Sellers et al. 1986; Sellers et al. 1996; Dickinson et al. 1993), slope hydrology (e.g., Liang et al. 1994; Takata et al. 2003), groundwater processes (e.g., Koirala et al. 2014) and vegetation dynamics (e.g., Cox et al. 2000; Montaldo et al. 2005; Ivanov et al. 2008). As the complexity of LSMs increases, they have many unknown soil and vegetation parameters which cannot be directly observed. Therefore, it is necessary to infer those unknown parameters from the observation data which include the information of model's state variables by evaluating gaps between simulation and observation. This parameter optimization is the grand challenge in hydrologic data assimilation.

There are two major obstacles in the parameter optimization of LSMs. First, model parameters nonlinearly affect simulated variables so that the cost function, which maps parameters to the gaps between simulation and observation, may have a lot of local minima. In addition, model parameters mutually interact with each other (e.g, one parameter is sensitive to simulated variables only if another parameter is above a threshold). Therefore, it is not straightforward even to just find the set of sensitive parameters to observable variables (e.g., Rosero et al. 2010; Pappas et al. 2013; Sawada and Koike 2014). Second, different combinations of parameters show similar skills to reproduce observations so that many combinations of parameters can be equally acceptable, which is called equifinality in hydrologic literature (Beven and Freer 2001). Searching a single optimal representation of parameters is clearly insufficient. A computationally efficient method to obtain the probability density function of parameters, which is called uncertainty assessment by some works in hydrology (e.g., Moradkhani et al. 2005), is crucially needed.

Despite a lot of efforts to optimize soil and vegetation parameters in LSMs, to our best knowledge, no study realized globally applicable parameter optimization and uncertainty assessment of LSMs by explicitly resolving the nonlinear effects of parameters and the equifinality. Towards a globally applicable method to optimize unknown parameters of LSMs, many previous studies have assimilated globally applicable satellite observations into LSMs. For example, Yang et al. (2007) developed the auto-calibration system, which can optimize soil parameters such as soil porosity, soil texture, and soil surface roughness, by assimilating microwave brightness temperature observed by Advanced Microwave Scanning Radiometer for Earth observation system (AMSR-E). Their system includes the Simple Biosphere model 2 (SiB2: Sellers et al. 1996) as an LSM, and a radiative transfer model to convert the LSM's state variables to microwave brightness temperature. Although they successfully obtained a single optimal set of the parameters using the Shuffled Complex Evolution (SCE) algorithm (Duan et al. 1992), they provided no uncertainty assessment (see also Yang et al. 2005; Yang et al. 2009; Qin et al. 2009; Tian et al. 2009). Sawada and Koike (2014) extended the work of Yang et al. (2007) by incorporating a dynamic vegetation model into the LSM. Since microwave brightness temperature is sensitive to both surface soil moisture and vegetation water content, simultaneous estimation of both hydrologic and ecologic model parameters was realized and the skill of the LSM to simultaneously simulate both soil moisture and vegetation dynamics was improved. However, no uncertainty assessment was provided in their work. Bandara et al. (2015) optimized the parameters of the soil property in the Joint UK Land Environment Simulator (JULES: Best et al. 2011) using a satellite-observed soil moisture product based on Soil Moisture and Ocean Salinity (SMOS) (see also Bandara et al. 2013). They successfully retrieved a single optimal soil property by the particle swarm optimization (Kennedy and Eberhart 1995) without the information of uncertainty. Kato et al. (2013) optimized 24 ecological parameters using satellite-observed fraction of absorbed photosynthetically active radiation. They realized the estimation of both optimal parameters and their uncertainties by a variational method although they assumed the Gaussian probability density distribution of parameter uncertainties and the linear relationship between parameters and observations.

Many previous hydrologic studies proposed methods based on the Markov Chain Monte Carlo (MCMC) sampler and successfully applied them to parameter optimization and uncertainty assessment of lumped hydrological models (e.g., Vrugt et al. 2003; Vrugt et al. 2008; Schoups and Vrugt 2010; Vrugt et al. 2013) and distributed LSMs (e.g., Vrugt et al. 2003; Rosero et al. 2010). The limitation of those MCMC-based algorithms is that many trials of model integration ($\mathcal{O}(10^3 \sim 10^6)$) are required and they cannot be completely parallelized. Therefore, it is infeasible to directly apply the previously proposed MCMC-based algorithms to the globally distributed parameter optimization and uncertainty assessment of LSMs considering their computational cost.

Here I aim to accelerate parameter optimization and uncertainty assessment of an LSM using the technique of statistical machine learning based surrogate modeling which is theoretically investigated in the field of applied mathematics called uncertainty quantification (Sullvan 2015). Although this technique has been used for the parameter sensitivity analysis of atmospheric models (e.g., Qian et al. 2018), hydrological models (e.g., Dell'Oca et al. 2017; Maina and Guadagnini 2018; Parente et al. 2019), and ecological models (e.g., Hawkins et al. 2019), few studies have applied it to parameter optimization and uncertainty assessment of LSMs with globally applicable satellite observations. In this study, a statistical surrogate model, which can mimic the relationship between model parameters and gaps between simulation and observation, is developed using machine learning. Then, using the statistical surrogate model, which is computationally cheaper than the original LSM, the probability distribution of parameters is sampled by the MCMC sampler. I demonstrate the potential of this approach in both synthetic and real-data experiments.

## 2. Methods
### 2.1. Problem statement
In this section, the problem of parameter optimization and uncertainty assessment is formulated. A discrete-time state-parameter dynamical system can be formulated as:
$$\boldsymbol{x}_{t+1} = M(\boldsymbol{x}_t, \boldsymbol{\theta}, \boldsymbol{u}_t) + \boldsymbol{q}_t \quad (1)$$
where $\boldsymbol{x}_t$ is the state variable (e.g., soil moisture, soil temperature and biomass) at time t, $M$ is the LSM, $\boldsymbol{\theta}$ is the model parameters, $\boldsymbol{u}_t$ is the external forcing (e.g., precipitation and incoming radiation), and $\boldsymbol{q}_t$ is the noise process which represents the model error. Since the whole space of the state variables cannot be observed, it is useful to formulate an observation process as follows:
$$\boldsymbol{y}_t^f = h(\boldsymbol{x}_t) + \boldsymbol{r}_t \quad (2)$$
where $\boldsymbol{y}_t^f$ is the simulated observation (e.g., microwave brightness temperature) at time t, $h$ is the observation operator which is a radiative transfer model in this study, and $\boldsymbol{r}_t$ is the noise process which represents the observation error.

Let,

$$y^f_{0:T} = \{y^f_0, y^f_1, ..., y^f_T\} \quad (3)$$
$$y^o_{0:T} = \{y^o_0, y^o_1, ..., y^o_T\} \quad (4)$$

where $y^o_t$ is the observation (e.g., microwave brightness temperature from a satellite) at time t. The purpose of parameter optimization and uncertainty assessment in this study is to obtain the posterior probability distribution of model parameters $\boldsymbol{\theta}$ by the timeseries of simulation (3) and observation (4) from time 0 to T and the Bayes' rule

$$p(\boldsymbol{\theta}|y^o_{0:T}) \propto p(y^o_{0:T}|\boldsymbol{\theta})p(\boldsymbol{\theta}) \quad (5)$$

where $p(\boldsymbol{\theta}|y^o_{0:T})$ is the posterior probability distribution of $\boldsymbol{\theta}$ given by the observations $y^o_{0:T}$, $p(y^o_{0:T}|\boldsymbol{\theta})$ is the likelihood, and $p(\boldsymbol{\theta})$ is the prior probability distribution of $\boldsymbol{\theta}$.

In the following section, I explain the LSM ($M$ in equation (1)), called EcoHydro-SiB, used in this study. The targeted model parameters $\boldsymbol{\theta}$ are also shown there. In section 2.3, I explain the observation operator ($h$ in equation (2)) used in this study. The method to efficiently obtain the posterior probability distribution of parameters ($p(\boldsymbol{\theta}|y^o_{0:T})$ in equation (5)) is explained in section 2.4.

## 2.2. Land surface model

EcoHydro-SiB, the improved version of SiB2 (Sellers et al. 1996), was developed as a module of a previously developed land data assimilation system, called Coupled Land and Vegetation Data Assimilation System (CLVDAS: Sawada and Koike 2014; Sawada et al 2015; Sawada 2018). In this section, the schemes related to parameter optimization and uncertainty assessment are explained. Please refer to Sawada and Koike (2014) for the complete description of this LSM. The values of parameters which are not optimized can also be found in Sawada and Koike (2014) and references therein. In this study, Ecohydro-SiB was run with the horizontal resolution of 0.25 degree and the timestep was 1 hour.

EcoHydro-SiB solves vertical interlayer water flows from surface to 2m depth using a one-dimensional Richards equation. The depth of surface soil layer was set to 5cm and that of the other soil layers was set to 10cm. To solve the Richards equation, capillary suction ψ (m) and hydraulic conductivity $K$ (m/s) are formulated as functions of volumetric soil moisture w (m$^3$/m$^3$) by the van Genuchten water retention curve (van Genuchten 1980):

$$\psi(w) = \frac{1}{\alpha}(S^{\frac{n}{n-1}})^{1/n} \quad (6)$$

$$\frac{K(w)}{K_s} = S^{1/2}[1 - (1 - S^{\frac{n}{n-1}})^{\frac{n-1}{n}}]^2 \quad (7)$$

$$S = \frac{w - w_r}{w_s - w_r} \quad (8)$$

where $K_s$ is the saturated hydraulic conductivity (m/s), $w_r$ is the residual water content (m³/m³), $w_s$ is the saturated water content or porocity (m³/m³), $\alpha$ and $n$ are the empirical parameters. In this study, $K_s$ and n are the targeted parameters to be optimized.

EcoHydro-SiB simulates vegetation growth and senescence by solving a carbon balance equation:

$$\frac{dC_{leaf}}{dt} = a_{leaf}NPP - (d_{leaf} + \gamma + \lambda)C_{leaf} \quad (9)$$

$$\frac{dC_{stem}}{dt} = a_{stem}NPP - d_{stem}C_{stem} \quad (10)$$

$$\frac{dC_{root}}{dt} = a_{root}NPP - d_{root}C_{root} \quad (11)$$

where $C_{leaf}$, $C_{stem}$, and $C_{root}$, are the carbon pools of leaves, stems, and roots, respectively (kg/m²), $a_{leaf}$, $a_{stem}$, and $a_{root}$ are the carbon allocation fractions of leaves, stems, and roots, respectively, $d_{leaf}$, $d_{stem}$, and $d_{root}$ are the normal turnover rates of leaves, stems, and roots, respectively, $\gamma$ and $\lambda$ are the stress factor related to water stress and temperature stress, respectively. See Sawada and Koike (2014) for the equations to calculate the allocation fractions and stress factors. NPP is the net primary production (mol m$^{-2}$ s$^{-1}$) and is calculated by the photosynthesis-conductance model. In this model, NPP is the function of the maximum Rubisco capacity of the top leaf, defined as $V_{max0}$ in this paper. See Sawada and Koike (2014) and Sellers et al. (1996) for details of the photosynthesis-conductance model. The linear relationship between the carbon pool of leaves and leaf area index (LAI) is assumed:

$$LAI = S_l C_{leaf} \quad (12)$$

where $S_l$ is the specific leaf area (m²/kg). Since stem and root biomass is required to support leaves, the following constraint is formulated:

$$C_{stem} + C_{root} \geq e_s C_{leaf} \quad (13)$$

where $e_s$ is the model parameter (see also Ivanov et al. 2008). If the relationship shown in equation (13) is violated, $a_{leaf}$ is set to zero and no carbon biomass is allocated to leaves. In this study, $V_{max0}$ and $e_s$ are the targeted parameters to be optimized.

## 2.3. Radiative transfer model

Note that any satellite retrieval products such as soil moisture and vegetation optical depth were not used in this study. Instead, I directly used brightness temperature data so that a nonlinear observation operator to convert the LSM's state variables to brightness temperature is needed. As an observation operator, I used the radiative transfer model which have already been installed in CLVDAS (Sawada and Koike 2014; Sawada et al 2015; Sawada 2018) to convert the model state variables into microwave brightness temperature. The radiative transfer model used in this study can explicitly simulate the dependence of microwave brightness temperature on both surface soil moisture and vegetation water content. The microwave radiative transfer in the vegetation canopy was simulated by the omega-tau model of Mo et al. (1980). Since this model explicitly calculates the emission and absorption by vegetation water, simulated microwave brightness temperature strongly depends on LAI calculated by equations (9)-(12). Land surface soil emissivity was calculated by the advanced integral equation model with the incorporation of a shadowing effect (Kuria et al. 2007). Since this model explicitly calculates the dielectric constant of the soil-water mixture, simulated land surface emissivity strongly depends on surface soil moisture. Please refer to Kuria et al. (2007) and Sawada and Koike (2014) for the complete description of the radiative transfer model.

Although there are some parameters in the radiative transfer model, no parameters were optimized in this study. Sawada et al. (2016) and Sawada et al. (2017) showed that the parameters of the radiative transfer model in the C- and X-band microwave can be accurately constrained by the in-situ observation experiment. The findings of these previous works (i.e., parameter values) have already been included in the radiative transfer model used in this study. I assumed that the uncertainty of the observation operator is significantly small compared with the LSM's uncertainty.

## 2.4. MCMC with a statistical surrogate model

As shown in section 2.2, there are four LSM's parameters to be optimized in this study: saturated hydraulic conductivity $K_s$, parameter of van Genuchten water retention curve $n$, maximum Rubisco capacity of the top leaf $V_{max0}$, and the factor controlling the relation between the carbon pools $e_s$. The four optimized parameters were normalized by the followings:

$$\frac{K_s}{K_s^{def}} = \theta_{1,min} + (\theta_{1,max} - \theta_{1,min})\theta_1 \quad (14)$$

$$\frac{n}{n^{def}} = \theta_{2,min} + (\theta_{2,max} - \theta_{2,min})\theta_2 \quad (15)$$

$$\frac{V_{max0}}{V_{max0}^{def}} = \theta_{3,min} + (\theta_{3,max} - \theta_{3,min})\theta_3 \quad (16)$$

$$\frac{e_s}{e_s^{def}} = \theta_{4,min} + (\theta_{4,max} - \theta_{4,min})\theta_4 \quad (17)$$

where $\theta_1$, $\theta_2$, $\theta_3$, and $\theta_4$ are the scaled parameters which range from 0 to 1, the superscript def means the "default" value for each parameter. The default values were specified according to the global soil and land use map (see section 3). For instance, $K_s$ can move from $\theta_{1,min} K_s^{def}$ ($\theta_1 = 0$) to $\theta_{1,max} K_s^{def}$ ($\theta_1 = 1$) in the optimization scheme. The possible ranges of parameters such as $\theta_{1,min}$ and $\theta_{1,max}$ are summarized in Table 1.

The prior $p(\boldsymbol{\theta})$ ($\boldsymbol{\theta} = [\theta_1, \theta_2, \theta_3, \theta_4]$) in equation (5) is assumed to be the bounded uniform distribution. Throughout the proposed algorithm shown below, parameters outside the range shown in Table 1 (see also equation (14)-(17)) were not sampled so that it is not possible to make the optimized parameters too far away from the initial default parameters, which prevents overfitting.

It should be noted that the selection of the prior $p(\boldsymbol{\theta})$, which implicitly includes the selection of the targeted parameters, the possible ranges of the parameters, and the shape of the distribution, was subjective (see also Pappas et al. 2013) although it was based on the previous works' experiences of parameter sensitivity analysis and parameter calibration. Sawada and Koike (2014) found that the parameters of $K_s$, $n$, and $V_{max0}$ are sensitive to simulated brightness temperature in the wide variety of climate conditions although the LSM used in Sawada and Koike (2014) did not include the equation (13) and $e_s$. Especially for vegetation parameters, I chose the parameters whose values have not been thoroughly discussed in the context of the LSM. I assumed that the "default" values of soil parameters (i.e., $K_s$ and $n$) were more accurate than those of vegetation parameters (i.e., $V_{max0}$ and $e_s$) because global datasets provided the different soil parameters in different grids using soil texture and empirical equations while vegetation parameters were assigned to the finite plant functional types. Therefore, I specified the larger ranges in comparison to their variances for vegetation parameters than soil parameters. The selection of the prior $p(\boldsymbol{\theta})$ significantly affects the results of parameter optimization and uncertainty assessment. However, in this study, I focused on the efficient parameter optimization and uncertainty assessment given the specific prior and the objective selection of the prior is not the scope of this study.

The posterior $p(\boldsymbol{\theta}|\boldsymbol{y}_{0:T}^o)$ in equation (5) was obtained by the MCMC sampler. I used the Metropolis-Hastings algorithm (Hastings 1970). First, the initial parameter vector $\boldsymbol{\theta}_0$ was specified. Then, the processes shown below were iterated:

1. For each iteration i, generate a candidate parameter vector $\boldsymbol{\theta}^c$. It is sampled from the distribution $q(\boldsymbol{\theta}^c|\boldsymbol{\theta}_i)$.
2. Evaluate the cost function $C(\boldsymbol{\theta}^c)$. The cost function C should be the indicator of the difference between simulation and observation. Larger C indicates that the gap between simulation and observation is smaller.
3. Calculate the acceptance ratio $a = \frac{C(\boldsymbol{\theta}^c)}{C(\boldsymbol{\theta}_i)}$.
4. Generate a random number b from the uniform distribution of [0,1]. Then,
    If $b \leq a$, accept the candidate parameter vector and $\boldsymbol{\theta}_{i+1} = \boldsymbol{\theta}^c$.
    If $b \geq a$, reject the candidate parameter vector and $\boldsymbol{\theta}_{i+1} = \boldsymbol{\theta}_i$

In this study, the distribution $q(\boldsymbol{\theta}^c|\boldsymbol{\theta}_i)$ was set to the Gaussian distribution whose mean and standard deviation were set to $\boldsymbol{\theta}_i$ and 0.1, respectively. The cost function C should be proportional to the targeted distribution $p(\boldsymbol{\theta}|\boldsymbol{y}_{0:T}^o)$. The cost function C was defined using the root-mean-square error (RMSE):

$$C = \exp\left(-\frac{RMSE}{\sigma_o}\right) \quad (18)$$

$$RMSE = \sqrt{\frac{1}{T}\sum_{t=0}^{T}(y_t^f - y_t^o)^2} \quad (19)$$

where $\sigma_o$ is the observation error in the satellite-observed brightness temperature and was set to 1 (K) in this study. See also equations (2)-(4).

It is infeasible to directly apply the MCMC sampler to parameter optimization and uncertainty assessment of the LSM because whenever the cost function C is evaluated, I need to run the LSM and the radiative transfer model for the long period which includes the period of spinup to obtain the simulated observation formulated in equation (3). The Metropolis-Hastings algorithm requires the $10^4$~$10^6$ iterations and these iterations cannot be parallelized. Therefore, the MCMC-based algorithms which are more efficient than the direct application of the Metropolis-Hastings algorithm have been proposed (e.g., Vrugt et al. 2003; Vrugt et al. 2008) and I used a statistical surrogate model.

First, I generated 400 ensemble members of parameters $\boldsymbol{\theta}$ by the Latin hypercube sampling (Deutsch and Deutsch 2012). The library used in this study to generate ensemble members can be found in https://github.com/sahilm89/lhsmdu. Second, I evaluated RMSE and the cost function C for 400 ensemble members by driving the LSM and the radiative transfer model and comparing simulation and observation (see equations (3), (4), (18), and (19) and refer to Section 3 for the details of the observation data). The dataset which includes 400 combinations of $\boldsymbol{\theta}$ and C was obtained. Third, by learning this dataset, a statistical machine learning based surrogate model, $g(\boldsymbol{\theta})$, was developed. I chose the Gaussian process regression (Rasmussen and Williams 2006) with the Matern kernel as a machine learning method. The python library, scikit-learn (https://scikit-learn.org/stable/modules/gaussian_process.html), was used. The input of g is only parameters $\boldsymbol{\theta}$ and the output of g is only the cost function C. The surrogate model generated in this study cannot calculate any real-world state variables such as soil moisture and LAI. Instead, the surrogate model g can mimic the response of $\boldsymbol{\theta}$ to the cost function C by the much cheaper computational cost than the full model (i.e. the LSM and the radiative transfer model). Fourth, I applied the MCMC sampler in which the cost function $C(\boldsymbol{\theta}^c)$ was evaluated by the surrogate model $g(\boldsymbol{\theta}^c)$.

To develop the statistical surrogate model, $g(\boldsymbol{\theta})$, RMSE was normalized:

$$RMSE_{norm} = \frac{RMSE - RMSE_{min}}{RMSE_{max} - RMSE_{min}} \quad (20)$$

where $RMSE_{max}$ and $RMSE_{min}$ are the maximum and minimum RMSEs in the 400 ensemble members, respectively. The Gaussian process regressor learned the relationship between $\boldsymbol{\theta}$ and $RMSE_{norm}$ and predicted $RMSE_{norm}$ as a function of $\boldsymbol{\theta}$. Then, $g(\boldsymbol{\theta})$ predicted the cost function:

$$g(\boldsymbol{\theta}) = \exp(-\frac{1}{\sigma_o}(\widehat{RMSE_{norm}}(\boldsymbol{\theta}) \times (RMSE_{max} - RMSE_{min}) + RMSE_{min})) \quad (21)$$

where $\widehat{RMSE_{norm}}(\boldsymbol{\theta})$ is the predicted normalized RMSE by the Gaussian process regression.

In summary, the proposed method is the following:

1. Generate 400 ensemble members of model parameters $\boldsymbol{\theta}$ by the Latin hypercube sampling.
2. Drive the LSM and the radiative transfer model and calculate RMSE using those parameter ensembles in parallel.
3. Construct the statistical surrogate model $g(\boldsymbol{\theta})$ from the dataset of the $\boldsymbol{\theta}$- C combinations by the Gaussian process regression.
4. Run the Metropolis-Hastings algorithm with $10^5$ iterations in which the cost function $C(\boldsymbol{\theta}^c)$ was evaluated by the statistical surrogate model $g(\boldsymbol{\theta}^c)$. Retrieve the posterior distribution $p(\boldsymbol{\theta}|\boldsymbol{y}^o_{0:T})$.

This optimization and uncertainty assessment method was applied for every model horizontal grid individually so that the different surrogate models were generated for different model horizontal grids in space.

## 3. Data

The International Satellite Land Surface Climatology Project 2 (ISLSCP II) data (Global Soil Data Task Group, 2000) and the Food and Agricultural Organization (FAO) global dataset (Food and Agricultural Organization 2003) were used to derive the LSM's default parameters (see also section 2.4).

The Global Land Data Assimilation System (GLDAS) v2.1 meteorological forcing (Rodell et al. 2004; Sheffield et al. 2006) was used to drive the LSM. The meteorological forcing data necessary to drive the LSM are surface pressure, precipitation, surface air temperature, relative humidity, incoming solar radiation, incoming longwave radiation, and wind speed. The original temporal resolution of this data is 3-hourly and the data were linearly interpolated to hourly, which is consistent to the timestep of the LSM (see section 2.2). The original horizontal resolution of this data is 0.25 degree and is consistent to the horizontal resolution of the LSM (see section 2.2).

The observed microwave brightness temperature is from the AMSR-E L3 product provided by Japan Aerospace eXploration Agency (JAXA) (Kachi et al. 2013). I resampled it from the native resolution to the 0.25 degree to make the horizontal resolution consistent to that of the LSM (see section 2.2). Brightness temperatures at 6.925 and 10.25 GHz were used for parameter optimization and uncertainty assessment since there were small atmospheric effects and the large sensitivity to the land conditions in these frequencies. Both horizontally and vertically polarized data were used. I used only night scene data to reduce the effects of surface temperature bias. The temporal resolution is approximately 2-daily.

The Global LAnd Surface Satellite LAI (GLASS LAI) product (Xiao et al. 2013) was used as independent data (i.e. data which were not used for parameter optimization) to validate the skill of the LSM with parameter optimization to simulate vegetation dynamics. The GLASS LAI was generated from MODerate resolution Imaging Spectroradiometer (MODIS) visible and infrared observations. The data were resampled from the native resolution to 0.25 degree which is consistent to the horizontal resolution of the LSM (see Section 2.2).

The European Space Agency Climate Change Initiative Soil Moisture (ESA CCI SM) v3.2 product (Dorigo et al. 2017) was used to validate the skill of the LSM to simulate surface soil moisture. It should be noted that ESA CCI SM is not completely independent to the AMSR-E since it includes the AMSR-E based soil moisture retrieval. However, many other satellite observations are also included in this dataset so that ESA CCI SM is useful as reliable surface soil moisture data. The spatial resolution of the data is 0.25 degree which is consistent to the LSM (see Section 2.2).

The Global Land Evaporation Amsterdam Model (GLEAM) v3 (Martens et al. 2017) was used to validate the skill of the LSM to simulate evapotranspiration. Note that the GLEAM's evapotranspiration is calculated by an LSM (not observation). Because ESA CCI SM is assimilated into the LSM in GLEAM, the data is also dependent to the AMSR-E dataset. The spatial resolution is 0.25 degree and the temporal resolution was daily.

## 4. Experiment Design
### 4.1. Study area
The study area is a part of the Sahel region and is shown in Figure 1. This study area was chosen because there was a steep gradient of climate and landscape from the rain forest in the southern part of the study area to the savanna and desert in the northern part of the study area. While we focus on a specific region in this study, the proposed method can be applicable to other climates and biomes. I could discuss how the proposed algorithm works in the different climate and vegetation conditions. The Sahel area has been chosen as study areas in the other land data assimilation studies (e.g., Albergel et al. 2018) by the same reason.

## 4.2. Synthetic experiment

To deeply discuss the advantages and limitations of the proposed algorithm, an ideallized synthetic experiment was performed. In the synthetic experiment, I specified the synthetic true model parameters, shown in Table 1, and drove the LSM to generate the synthetic true land state. From the synthetic true land state, the synthetic observation (i.e. microwave brightness temperature at 6.925 and 10.65 GHz) was generated using the radiative transfer model. Using this simulated observation by the synthetic true parameters, the algorithm described in section 2.4 was performed. The timing of the synthetic observation was the same as real observations from AMSR-E (see Section 3). The meteorological forcing and the default parameter values were retrieved from the real data described in Section 3. Since many unknown sources of uncertainty exist in real— data experiments, the synthetic experiment is crucial to eliminate them and verify the newly-developed algorithm. In addition, while it is extremely difficult to obtain "true" model parameters from the real data, the synthetic experiment makes it easier to interpret the results by comparing the estimated and synthetic true parameters. The synthetic experiment has been widely adopted in the atmospheric and hydrologic communities (e.g., Zhang et al. 2016; Moradkhani et al. 2005; Vrugt et al. 2013; Bandara et al. 2013; Abolafia-Rosenzweig et al. 2019).

This synthetic experiment was performed in the three points, the Site I-III shown in Figure 1b, which have different climatic and vegetation conditions. The study period was 2003-2010 which is almost identical to the AMSR-E's operational period. Whenever I drove the LSM including the generation of the synthetic truth, the model integration from 2003 to 2010 was repeated 4 times in order to spin up the initial condition of state values and then the additional model integration from 2003 to 2010 was used for the final results of the simulation.

## 4.3. Real-data experiment

In addition to the idealized synthetic experiment, I demonstrated the potential of the proposed algorithm in the real-world application. Here I used the real AMSR-E observed brightness temperature data for parameter optimization and uncertainty assessment. The study area was shown in Figure 1b. The study period and the method to spinup the model were identical to the synthetic experiment (see section 4.2). Although the "true" parameters cannot be known in the real-world application, the performance of the proposed algorithm was evaluated by the independent satellite data (see section 3). In addition, I demonstrated how the proposed algorithm contributes to the identification of sensitive parameters and equifinality.

In this main manuscript, the AMSR-E data from 2003 to 2010 were used for parameter optimization and uncertainty assessment and I evaluated the performance of the LSM from 2003 to 2010. In this case, the training phase is identical to the verification phase (note that the training data (i.e. AMSR-E brightness temperature) are different from the verification data (i.e. GLASS LAI, ESA CCI, and GLEAM)). I also performed the experiment in which the training phase is not identical to the verification phase. In this experiment, the parameter optimization and uncertainty assessment were performed by the AMSR-E data from 2003 to 2005. Then, the performance of the LSM was evaluated using the data from 2006 to 2010. The results of this experiment can be found in the supplement material.

## 5. Results
### 5.1. Synthetic experiment

All optimization procedures, shown in section 2.4, can be done in less than 10 minutes for a single model grid on 7 nodes (476 cores) of the supercomputer OakForest-PACS (https://www.cc.u-tokyo.ac.jp/en/supercomputer/ofp/service/). The total wall-clock time which includes the generation of the surrogate model and MCMC with the surrogate model is approximately 2 minutes so that the process of machine learning is computationally cheap. Although it took approximately 5 minutes to run the LSM and the radiative transfer model for the 40 years integration (note that it includes the model spin-up) by a single core, the advantage of the proposed method is that all of the required model runs (i.e. 400 ensemble runs) can be done completely in parallel. On the other hand, many model runs could not be parallelized if I directly applied the Metropolis-Hastings algorithm to the full model (i.e., the LSM and the radiative transfer model). Total computation time will be approximately $5 \times 10^5$ minutes when the Metropolis-Hasting algorithm with $10^5$ iterations is applied to the full model. Therefore, the significant reduction of the computational time was achieved by the statistical machine learning based surrogate modeling.

In the Site I (0.125E, 10.125N see Figure 1b), annual precipitation is approximately 1190 (mm/yr). In the synthetic truth, LAI varies temporally between 0.5 and 2 so that the site is moderately vegetated and surface soil moisture can be observed from the satellite especially in the dry seasons (see Sawada et al. 2017 for the sensitivity of microwave brightness temperature to surface soil moisture in the vegetated area).

Figure 2a shows the response of each parameter to the gap between simulation and observation. For instance, blue dots show RMSEs of brightness temperature between the synthetic truth and the model simulation in which the normalized parameters of hydraulic conductivity ($\theta_1$ in equation (14)) were randomly drawn from 0 to 1, but the other parameters ($\theta_2, \theta_3$, and $\theta_4$ in equations (15)-(17)) were fixed to the synthetic truth shown in Table 1. Because the synthetic truth of normalized saturated hydraulic conductivity $\theta_1$ is set to 0.75 (see Table 1), RMSE is small when $\theta_1$ is around 0.75. As $\theta_1$ deviates from 0.75, RMSE increases. Figure 2a indicates that the response of parameters to RMSE is mostly linear except for n in the van Genutchen water retention curve (orange dots in Figure 2a).

As described in section 2.4, I drove the LSM with the 400 ensembles of the parameters randomly drawn by the Latin hypercube sampling. Then the statistical machine learning based surrogate model was constructed by the Gaussian process regression to predict RMSE for the new parameters which is not included in the training dataset. Here I generated the other 400 ensembles and drove the LSM and the radiative transfer model with them to provide the validation dataset of the $\boldsymbol{\theta} - RMSE$ relationship. Figure 3a indicates that the Gaussian process regressor accurately predicts RMSEs simulated by the LSM and the radiative transfer model ($R^2$=0.97). Therefore, the constructed statistical surrogate model can accurately mimic the response of the parameters to the gap between simulation and observation and it can be justified to use this surrogate model for parameter optimization and uncertainty assessment.

Figures 4a, 4b, 4c, and 4d show the posterior distribution sampled by the $10^5$ iterations of the Metropolis-Hastings algorithm with the statistical machine learning based surrogate model. The proposed algorithm can successfully retrieve the non-parametric posterior probabilistic distributions of the unknown parameters by the cheap computational cost. The modes of the distributions of saturated hydraulic conductivity ($\theta_1$), n of the van Genutchen water retention curve ($\theta_2$), and maximum rubisco capacity at top leaf ($\theta_3$) are reasonably consistent to the synthetic truth (Figures 4a, 4b, and 4c) while that of the factor

controlling the relation between the carbon pools ($\theta_4$) is not the case (Figure 4d). This is because the equifinality exists between $\theta_3$ and $\theta_4$. Figure 4e shows the joint distributions of two selected parameters. Among the sampled parameters, $\theta_3$ and $\theta_4$ are strongly correlated (slope = 0.689, R = 0.598 (p<0.001)). Many different combinations of $\theta_3$ and $\theta_4$ have similar performances to reproduce observed microwave brightness temperature. Large maximum rubisco capacity at top leaf ($\theta_3$) and small factor controlling the relation between the carbon pools ($\theta_4$) have the similar contribution to the LSM's state variables (i.e. increasing biomass allocated to leaves). In other words, the current observation has the insufficient information to perfectly constrain those parameters. It should be noted that the posterior distributions do not follow the Gaussian distribution, which shows that the non-parametric approach by MCMC is appropriate.

I drove the LSM with the 400 ensembles of the parameters drawn from the posterior distribution shown in Figures 4a-4d. I compared this simulation driven by the parameters from the posterior, $p(\boldsymbol{\theta}|\boldsymbol{y}_{0:T}^o)$ in equation (5), with the simulation driven by the parameters from the prior, $p(\boldsymbol{\theta})$ in equation (5) (i.e. random draws from the uniform distribution [0,1]). Figure 5 shows that the parameter optimization successfully improves the skill of the LSM to simulate observable microwave brightness temperature.

Figure 6 indicates that the parameter optimization also improves the skill of the LSM to reproduce the synthetic truth of LAI and soil moisture. The parameter optimization substantially reduces the uncertainty in the simulation of LAI and the median of LAI simulated by the MCMC-sampled parameters was more consistent to the synthetic truth than the simulation by the completely random samples of the parameters (Figures 6a and 6b). The parameter optimization also reduces the uncertainty in the simulation of surface soil moisture especially in the dry seasons (Figures 6c and 6d). Although the root-zone soil moisture cannot be directly observed by the satellite, the parameter optimization positively impacts to the simulation of root-zone soil moisture (Figure 6e and 6f) because the parameter adjustment affects the whole state space of the LSM.

In the Site II (0.625E, 7.125N see Figure 1b), annual precipitation is approximately 1590 (mm/yr). In the synthetic truth, LAI varies temporally between 2 and 4. In this densely vegetated condition, microwave emission from land soil surface is strongly attenuated by canopy and microwave brightness temperature is always insensitive to surface soil moisture (Sawada et al. 2017) so that the observation has the information of only vegetation dynamics in the whole period. In addition, the contribution of vegetation water content to microwave radiative transfer becomes saturated when LAI is large (see Sawada et al. 2017). Therefore, the observation is less informative in the Site II than the Site I.

Figure 2b indicates that the response of the parameters to RMSE in the Site II is similar to the Site I. Figure 3b indicates that the Gaussian process regressor accurately predicts the RMSEs calculated by the LSM and the radiative transfer model in the Site II ($R^2=0.99$) as shown in the Site I.

However, the posterior distributions of the parameters recover the synthetic truth less accurately than the Site I. Figures 7a, 7b, 7c, and 7d show that the posterior distributions of saturated hydraulic conductivity ($\theta_1$), n of the van Genutchen water retention curve ($\theta_2$), and maximum rubisco capacity at top leaf ($\theta_3$) in the Site II is flatter than the Site I and their mode is not consistent to the synthetic truth (red dashed lines). Although the sampled parameters successfully minimize RMSE between simulated and observed brightness temperatures (Figure 8), the observation is not very informative to constrain the four unknown parameters in the densely vegetated area. Figure 7e shows that there are correlations between sampled parameters such as $\theta_1$-$\theta_2$, $\theta_2$-$\theta_3$, and $\theta_3$-$\theta_4$. The advantage of the proposed method is that the equifinality due to the insufficiently informative observation can be explicitly detected by the relatively cheap computational cost.

The parameters drawn from the posterior distributions substantially reduce the uncertainty in the simulation of LAI (Figures 9a and 9b). Figure 9b shows that the simulation with the optimized parameters still has the relatively large uncertainty in the seasonal peak of LAI. This is because the contribution of vegetation water content to microwave radiative transfer becomes saturated when LAI is large (e.g., LAIs of 3.5 and 4.5 cannot be accurately distinguished by the microwave signal; see Sawada et al. 2017). Figures 9c, 9d, 9e, and 9f indicate that the optimization has a smaller impact on the soil moisture simulation than the Site I.

In the Site III (0.125E, 20.125N), annual precipitation is approximately 110 (mm/yr). This virtual site is located in the arid area and there is no vegetation in the synthetic truth.

Figure 2c indicates that the optimization problem in the Site III is completely different from those in the Sites I and II. Parameters of saturated hydraulic conductivity ($\theta_1$), maximum rubisco capacity at top leaf ($\theta_3$), and the factor controlling the relation between the carbon pools ($\theta_4$) are not sensitive to RMSE and the gap between simulation and observation is mainly controlled by the n of the van Genuchten water retention curve ($\theta_2$). Moreover, the response of $\theta_2$ to RMSE is quite non-linear. RMSE suddenly increases when $\theta^2$ is larger than ~0.5.

The Gaussian process regressor is less accurate to predict RMSE calculated by the LSM and the radiative transfer model than the cases of the Sites I and II (Figure 3c) although the predicted and true RMSEs are reasonably correlated ($R^2=0.89$). Since the parameter range of the transition from low RMSE to high RMSE is small (Figure 2c), more complex statistical models and/or more efficient ensemble sampling strategies might be needed to resolve the nonlinear response of the parameters.

The proposed method accurately samples the posterior distribution even in this extreme case. Figures 10a, 10c, and 10d show that the parameters of saturated hydraulic conductivity ($\theta_1$), maximum rubisco capacity at top leaf ($\theta_3$), and the factor controlling the relation between the carbon pools ($\theta_4$) are uniformly distributed in the range [0,1], which implies little sensitivity of these parameters to microwave brightness temperature. Figures 10b shows the n of the van Genuchten water retention curve ($\theta_2$) is highly sensitive to microwave brightness temperature. This sensitivity analysis is consistent to the results shown in Figure 2c. The mode of the posterior distribution of $\theta_2$ deviates from the synthetic truth due to the small sensitivity of the change in $\theta_2$ from 0.2 to 0.5 (see the flat orange line in this range in Figure 2c). Figure 11 shows that this parameter optimization successfully minimizes the gap between simulated and observed brightness temperatures.

Figures 12c and 12d indicate that the parameter optimization significantly reduces the uncertainty in surface soil moisture. However, the skill to simulate sub-surface (35cm-45cm) soil moisture is degraded, which is not consistent to the results obtained in the Sites I and II. Although there is the sensitivity to sub-surface soil moisture in the range of the sampled parameters, all of the 400 ensembles simulate the almost identical brightness temperatures (Figure 11), LAI (Figure 12b) and surface soil moisture (Figure 12d). Therefore, the optimization reveals that the observation has the insufficient information to constrain sub-surface soil moisture in this site. Generally, in arid regions, surface variables (surface soil moisture and vegetation) are not well correlated with sub-surface soil moisture (e.g., Kumar et al. 2009) so that it is difficult to infer sub-surface soil moisture from surface observations.

The number of ensemble members strongly affects the performance of the surrogate model and the parameter optimization. The larger number of ensemble members can develop the more accurate surrogate model and sample the more accurate posterior distributions of unknown parameters. However, the larger ensemble size requires the larger computational resources, so that I need to find the reasonable ensemble size in order to accurately optimize the unknown parameters with the relatively cheap computational cost. To evaluate the sensitivity of the ensemble size, I performed the MCMC sampler with the surrogate model generated by the smaller number of ensemble members (50, 100, 200, and 300 members). Then, I compared the sampled posterior distributions by the smaller ensemble sizes with those shown in Figures 4, 7, and 10. The Kullback-Leiblur divergence (KLD) was used to measure the difference between two posterior distributions:

$$D_{KL}(p,q) = \sum_i p(i) log \frac{p(i)}{q(i)} \quad (22)$$

where $D_{KL}(p,q)$ is the KLD between two probabilistic distributions, p, and q. If two distributions are equal for all $i$, $D_{KL}(p,q) = 0$. If two distributions greatly differ from each other, the large value of $D_{KL}(p,q)$ is obtained. Therefore, the KLD can be used as an index to evaluate the closeness of the two probabilistic distributions. Here p is the posterior distributions generated by the surrogate model with the smaller ensemble size and q is that with the 400-ensemble size shown in Figures 4, 7, and 10.

Figure 13 shows the KLD as a function of ensemble size in the Sites I, II, and III. In all three sites, there are no large differences between sampled posteriors by 300-ensemble and 400-ensemble so that I cannot expect that the parameter optimization shown above is significantly improved by increasing the ensemble size from 400 members. Figure 13 also indicates that the surrogate model is not accurate enough to obtain the posterior of the unknown parameters if the ensemble size is less than 200. I also found that the skill of the surrogate model to estimate RMSE between simulated and observed brightness temperature becomes poor if I made the ensemble size less than 200 (not shown). It should be noted that this required ensemble size strongly depends on the number of the unknown parameters. If the number of the unknown parameters increases, larger ensemble size is required.

## 5.2. Real data experiment

Here I demonstrate the advantages and limitations of the proposed method in the real data experiment in which the real AMSR-E brightness temperature observations are assimilated into the LSM. While in the idealized synthetic experiment I assumed that the 4 selected unknown parameters are the sole source of errors, in the real data experiment I neglected many other error sources such as the parameters which are not optimized, the meteorological forcing ($\boldsymbol{u}_t$ in equation (1)), the model structure ($\boldsymbol{q}_t$ in equation (1)), and the observation process ($\boldsymbol{r}_t$ in equation (2)). This point should be noted for the interpretation of the results.

Figures 14a, 14b, 14c, and 14d show the medians of the $10^5$ sampled parameters of saturated hydraulic conductivity ($\theta_1$), n of the van Genutchen water retention curve ($\theta_2$), and maximum rubisco capacity at top leaf ($\theta_3$), and the factor controlling the relation between the carbon pools ($\theta_4$), respectively. The proposed method can obtain the spatially distributed optimal parameters by the reasonably cheap computational cost. Compared with the other parameters, the saturated hydraulic conductivity $K_s$ prefers the default variable (Figure 14a). Figure 14b shows that the optimal values of the van Genuchten's n tend to be smaller than the default values in the wetter area while they tend to be larger than the default values in the drier area. Figures 14c and 14d shows that the optimal values of maximum Rubisco capacity of the top leaf $V_{max0}$ and the factor controlling the relation between the carbon pools $e_s$ tend to be larger and smaller than the default values, respectively.

As shown in the synthetic experiment of the virtual Site III, the posterior distribution of the parameters with no significant sensitivity to observation tends to be a uniform distribution (see Figure 2c and Figures 10a, 10c, and 10d). Therefore, the Kullback-Leibler divergence (KLD) (Kullback and Leibler 1951) between the uniform distribution and the posterior distribution can be a good index of the parameter sensitivity. Large KLD means that the sampled posterior distribution of the parameters substantially differs from the uniform distribution and the parameter is sensitive to observations.

Figures 14e, 14f, 14g, and 14h show the KLDs of the four parameters. The saturated hydraulic conductivity $K_s$ tends to be uniformly drawn and less sensitive to observation than the other parameters (Figure 14e). Figure 14f shows that the van Genuchten's n is sensitive in the whole study area. This is because the van Genuchten's n affects not only soil water dynamics but also the soil moisture thresholds in vegetation water stress (i.e. field capacity and wilting point), which is crucially important to the simulation of vegetation growth and senescence. Figures 14g and 14h show that the maximum Rubisco capacity of the top leaf $V_{max0}$ and the factor controlling the relation between the carbon pools $e_s$ are sensitive only in the vegetated area (see also the distribution of observed LAI in Figure 1b).

Figure 15 shows the correlations between the sampled parameters. In the vegetated area, there are distinct correlations between different parameters as I showed in the Site II of the synthetic experiment (see Figure 7e). On the other hand, in the northern arid area, no such clear and consistent patterns of correlations can be found. This is because a single parameter (the van Genutchen's n) can be constrained by the observations in this region and the others cannot be constrained (see Figures 14e-h), which is consistent to the findings of the synthetic experiment in the Site III (see Figure 10e).

As I did in the section 5.1, I drove the LSM with the 400 ensembles of the parameters drawn from the posterior distribution. I compared this simulation driven by the parameters from the posterior, $p(\boldsymbol{\theta}|\boldsymbol{y}^o_{0:T})$ in equation (5), with the simulation driven by the parameters from the prior, $p(\boldsymbol{\theta})$ in equation (5) (i.e. random draws from the uniform distribution [0,1]). Unlike the idealized synthetic experiment, no truth of the LSM's state variables can be obtained. Instead, I compared the simulation with the independent observation data which are GLASS LAI, ESA CCI surface soil moisture, and GLEAM evapotranspiration (see section 3).

It is not straightforward to evaluate the ensemble simulation. Here I defined the two indices to evaluate the skill of the LSM with 400 parameter ensembles to reproduce observed LAI and surface soil moisture. First, I calculated the bias for each ensemble member and took its root-mean-square:

$$BIAS_{ens} = \sqrt{\frac{1}{N}\sum_{i=1}^{N}\overline{(y_i^f - y^o)}^2} \qquad (22)$$

where $N$ is the ensemble size, $y_i^f$ is the simulated observation by the ensemble $i$, $y^o$ is the observation, and $\overline{y_i^f - y^o}$ is the temporal mean of the simulation-observation differences, which is bias for the ensemble $i$. Second, I calculated the unbiased root-mean-square error (ubRMSE) by removing the bias between observation and ensemble mean.

$$ubRMSE_{ens} = \frac{1}{N}\sum_{i=1}^{N} ubRMSE_i \quad (23)$$

$$ubRMSE_i = \sqrt{E[((y_i^f - E(y^f) - (y^o - E(y^o))^2]} \quad (24)$$

where E( ) is the expectation operator and $E(y^f)$ is calculated as the temporally averaged ensemble mean. Finally, I calculated the improvement rate for each score (i.e., $BIAS_{ens}$ and $ubRMSE_{ens}$) as follows:

$$I_R = \frac{S_{mcmc} - S_{unif}}{S_{unif}}$$

where $I_R$ is the improvement rate, $S$ is the evaluation metrics ($BIAS_{ens}$ or $ubRMSE_{ens}$), the subscript mcmc means the evaluation metrics from the simulation with the MCMC sampled parameters, and the subscript unif means the evaluation metrics from the simulation with the parameters sampled from the prior uniform distribution.

Figures 16a and 16b show the spatial distribution of the improvement rates for $BIAS_{ens}$ and $ubRMSE_{ens}$ of LAI against GLASS LAI (see section 3), respectively. Although the $BIAS_{ens}$ was reduced by the parameter optimization in the large portion of the study area, there is the substantial degradation in the west part of 10N-12N. In this area, the optimized parameters $\theta_1$ and $\theta_2$ are unnaturally small (Figures 13a and 13b) due probably to the overfitting induced by the neglected errors such as errors in the model structure, the radiative transfer model, and the meteorological forcing. This overfitting might degrade the skill of the LSM to simulate LAI. Although the impact of the parameter optimization on $ubRMSE_{ens}$ of LAI is relatively small, there is the degradation in the part of the vegetated area. In this area, the seasonal cycle of the simulated LAI is slightly delayed to that of observed LAI (not shown), which can also be found in the other LSMs (e.g., Jarlan et al. 2008). This LSM's structural error cannot be fixed in the current parameter optimization framework.

Figures 16c and 16d show the spatial distribution of the improvement rates for $BIAS_{ens}$ and $ubRMSE_{ens}$ of surface soil moisture against ESA CCI (see section 3), respectively. The parameter optimization greatly increases $BIAS_{ens}$ in the northern arid area. However, $BIAS_{ens}$ in this area comes mainly from the gap between simulated and observed minimum baseline surface soil moisture in the dry seasons (not shown).

Considering the error in the satellite soil moisture retrievals (approximately 0.05 m$^3$/m$^3$) and the small impacts of this baseline soil moisture on the water, carbon, and energy balances, this gap should be removed for evaluation (e.g., Entekhabi et al. 2010). Figure 15d shows that the parameter optimization reduces $ubRMSE_{ens}$ more than 10 % in many parts of the study area.

Figures 16e and 16f show the spatial distribution of the improvement rates for $BIAS_{ens}$ and $ubRMSE_{ens}$ of evapotranspiration against GLEAM (see section 3), respectively. The impact of the proposed method on the simulation of evapotranspiration is similar to that on the simulation of surface soil moisture.

In the experiment described above, the parameter values are optimized by the 8-year (2003-2010) AMSR-E data. I found that the optimized parameter values and the correlations between them do not greatly change if the training period is shortened to 3 years (Figures S1 and S2). Figure S3 shows that the performance of the LSM was unchanged by splitting the study period into the 3-year training period and the 5-year validation period.

## 6. Discussions

In this study, the globally applicable method for parameter optimization and uncertainty assessment of the LSM was developed. Here I discussed the characteristics of the proposed method comparing with the existing methods.

First, the proposed method can retrieve the non-parametric posterior probabilistic distributions of the unknown parameters. Results in the synthetic experiment show that the posterior distribution of the unknown parameters given the passive microwave brightness observations follows the non-Gaussian probabilistic distribution. From the posterior distribution, I can quantify the complex effects of the multiple parameters on observable values (i.e. equifinality). It is difficult to perform the uncertainty quantification of the LSM's unknown parameters shown in this study if the optimization methods search a single set of parameters (e.g., Yang et al 2007; Sawada and Koike 2014) and assume that the posterior distribution follows the Gaussian distribution (e.g., Kato et al. 2013).

Second, the reasonably cheap computational cost was achieved by the technique of the statistical machine learning based surrogate modeling, which makes the proposed method globally applicable. The computational cost was largely determined by the number of iterations of the long-term LSM's integration to evaluate the long-term (more than 1-year) difference between simulation and observation. Despite a lot of previous efforts to reduce the number of iterations, many repetitions of the LSM's integration are still needed in the previously developed methods. For instance, Kato et al. (2013) efficiently minimized the cost function using the first and second derivatives of the cost function with respect to parameters. However, the optimization took approximately 40 iterations (see also Kaminski et al. 2012), which prevents the global application especially in the case that the long spin-up period is required and/or the long timeseries of simulation and observation need to be evaluated. Compared with the previous studies, the advantage of the proposed method is that no iterations of the LSM's integration are needed. It should be noted that the efficiency of the sequential data assimilation such as ensemble Kalman filtering (e.g., Hendricks-Franssen and Kinzelbach 2008) and particle filtering (e.g., Moradkhani et al. 2005), in which parameters are treated as elements of a state vector, may be equal to or greater than the proposed method since the repetition of the long-term simulation is not needed. However, these filters sequentially optimize parameters based on the short-term (less than a week) timeseries allowing parameters to temporally vary in the study period so that the problem which these filters solve is different from the problem stated in this study. Vrugt et al. (2013) found that the parameter optimization by particle filtering demonstrates non-stationary of the optimized parameters (i.e. posterior distributions are not converged) due to temporal changes in parameter sensitivity (i.e., parameters affects the model's dynamics differently in the different seasons), which is the potential limitation of these filters compared with the proposed method in this study.

Third, I effectively used the satellite observation of microwave brightness temperature which is sensitive to the important ecohydrological variables (i.e., soil moisture and vegetation water content) and has the global and all-weather capability. It also contributes to making the proposed method globally applicable. In this study, the observations from AMSR-series satellites were focused on because their observations have a strong and consistent sensitivity to both surface soil moisture and vegetation dynamics (Sawada et al. 2017). The proposed method can be easily extended to use the observations of L-band passive microwave from SMOS and Soil Moisture Active Passive (SMAP), which is more transparent to vegetation water content and has a deeper sensing depth of soil moisture (e.g., Chan et al. 2016).

## 7. Conclusion

In this study, I proposed the globally applicable and computationally efficient method for parameter optimization and uncertainty assessment of the LSM by combining MCMC with machine learning. In the idealized synthetic experiment and the real-data experiment, the proposed method successfully obtained the non-parametric posterior distribution of four unknown parameters and improved the skill of the LSM to simulate both soil moisture and vegetation dynamics. The advantage of the proposed method is its computational efficiency. The proposed method is 50,000 times as fast as the direct application of MCMC to the full LSM and more efficient than the other efficient algorithms proposed in the previous studies.

Towards the global-scale parameter optimization and uncertainty assessment of the LSMs, several challenges remain. First, the prior should be carefully designed. The prior stated in this study implicitly includes the selection of parameters, the range of parameters, and the shape of distributions. In addition, the explicit consideration of the error in meteorological forcing and model structure is helpful to further improve the performance of the LSM's simulation. It can be done using more than one dataset of meteorological forcing and multimodel ensembles. Second, in addition to the microwave satellite observations such as AMSR-E, there are many other important satellite observations which can contribute to parameter optimization and uncertainty assessment. Extending the framework proposed in this study to explicitly consider a number of observation types (i.e. a number of targets or objectives) is an important research topic.

**Acknowledgements**

This work was supported by the Japan Society for the Promotion of Science KAKENHI grant JP17K18352 and JP18H03800, the JAXA grant ER2GWF102, and the JST AIP Grant Number JPMJCR19U2. The ISLSCP II dataset can be downloaded at https://daac.ornl.gov/. The FAO dataset can be downloaded at http://www.fao.org/home/en/. The GLDAS forcing dataset was provided by the National Aeronautics and Space Administration (NASA) and can be downloaded at https://disc.gsfc.nasa.gov/. The AMSR-E dataset was provided by JAXA and can be downloaded at https://gportal.jaxa.jp/gpr/?lang=en. The GLASS LAI dataset can be downloaded at http://www.chinageoss.org/gee/2013/en/en1/en_A_2/. The ESA CCI soil moisture dataset can be downloaded at http://cci.esa.int/. I thank two anonymous reviewers for their constructive comments.

and prediction of tropical cyclones. 1–10. https://doi.org/10.1002/2016GL068468.

Table 1. Configurations of optimized parameters, their ranges, and the synthetic truth. See sections 2.4 and 4.2 for details.

| | descriptions | $\theta_{min}$ | $\theta_{max}$ | Truth |
|---|---|---|---|---|
| $\theta_1$ | Normalized parameter for saturated hydraulic conductivity $K_s$ | 0.5 | 1.5 | 0.75 |
| $\theta_2$ | Normalized parameter for van Genutchen water retention curve $n$ | 0.8 | 1.2 | 0.4 |
| $\theta_3$ | Normalized parameter for maximum Rubisco capacity of the top leaf $V_{max0}$ | 0.5 | 1.5 | 0.25 |
| $\theta_4$ | Normalized parameter for the factor controlling the relation between the carbon pools $e_s$ | 0.25 | 1.75 | 0.6 |

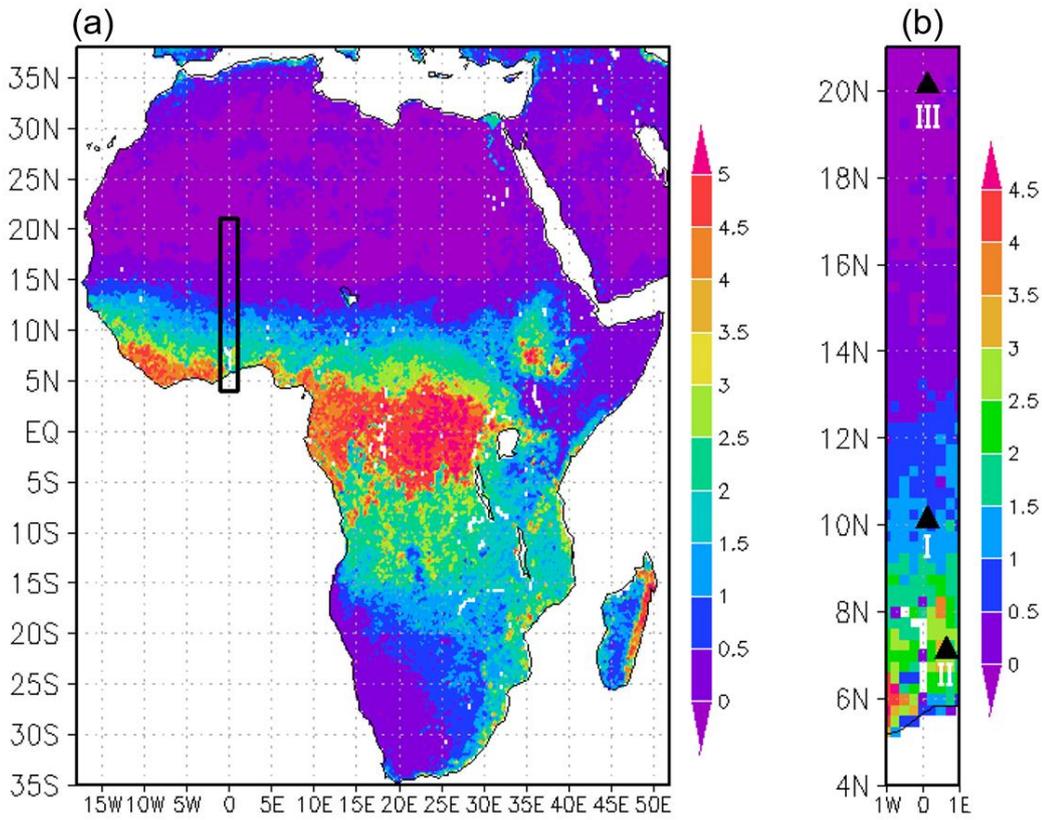

**Figure 1.** Study area. Yearly LAI from the GLASS LAI product is shown as colored shades. The area shown in (b) corresponds to the black box in (a) and indicates the study area of the real data experiment. Black triangles in (b) are the points in which the idealized synthetic experiments are implemented. See also Section 4.

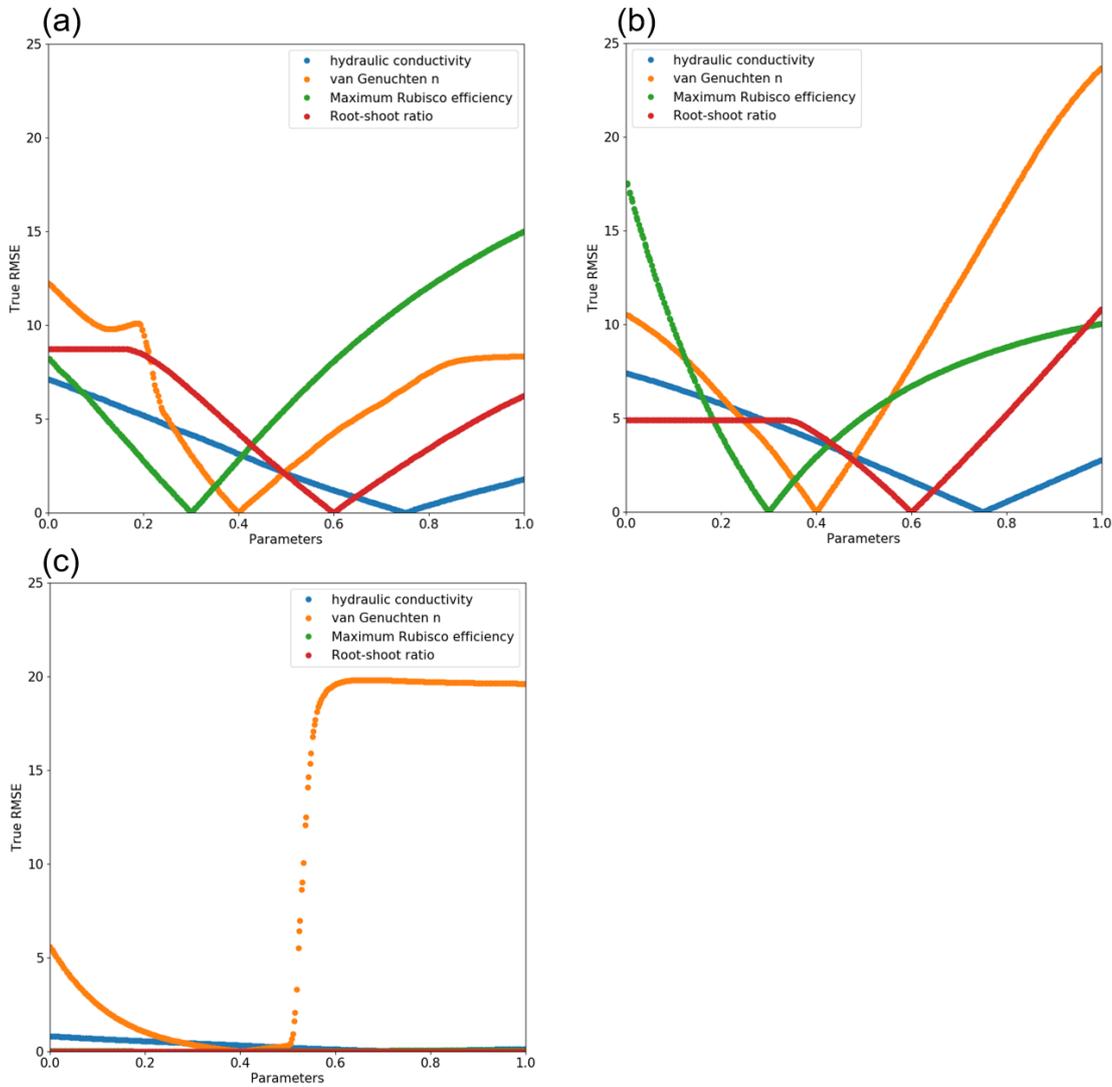

**Figure 2.** Simulated RMSEs of brightness temperature between the LSM with the synthetic true parameters and that with the random draws of the normalized parameters of saturated hydraulic conductivity (blue), the van Genuchten's n (orange), maximum Rubisco efficiency at the leaf top (green), and the factor controlling the relation between the carbon pools (red) in (a) the Site I, (b) the Site II, and (c) the Site III. In each dot, the single targeted parameter was randomly drawn, and the other parameters were fixed to the synthetic truth (see Table 1 and section 5.1).

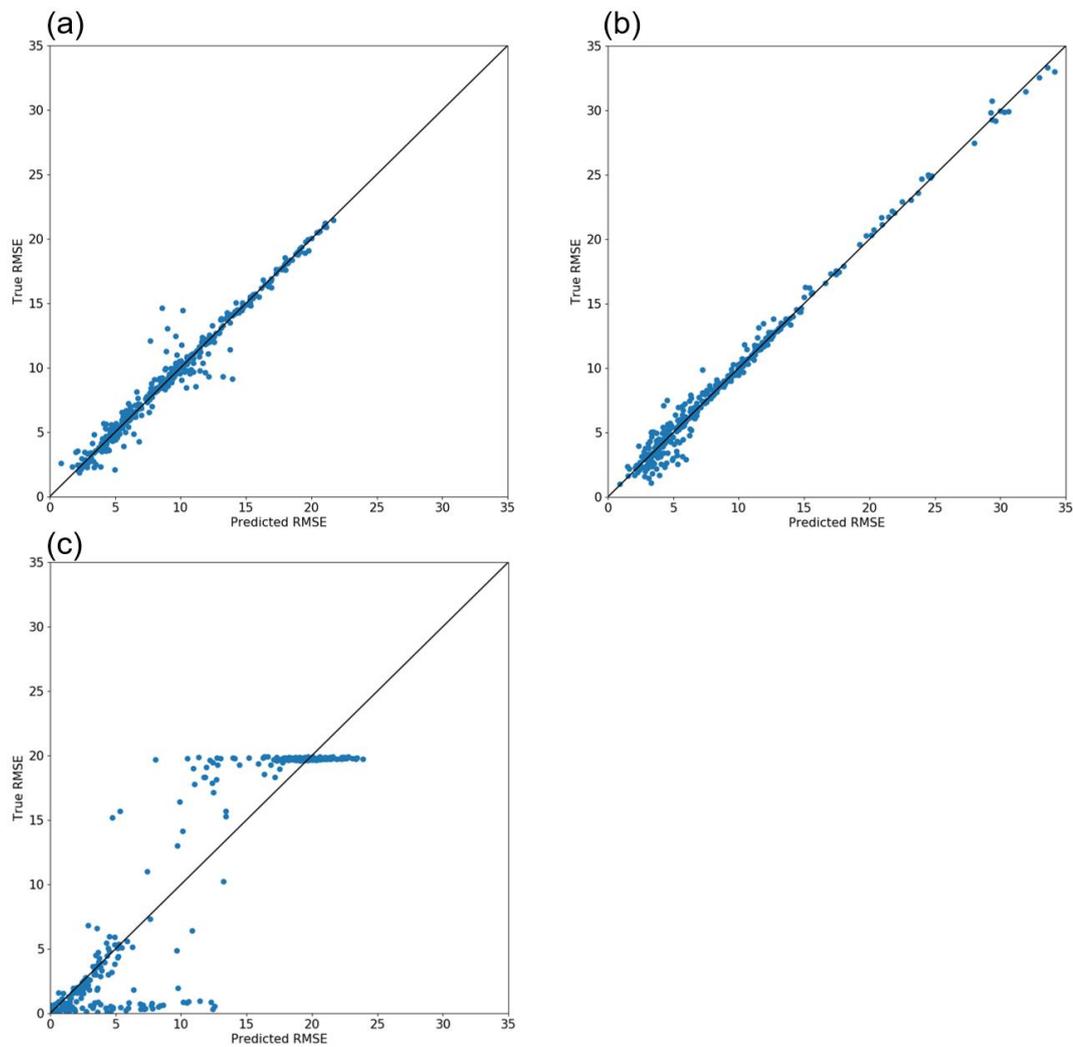

**Figure 3.** Comparisons between the synthetic true RMSEs calculated by the LSM and the radiative transfer model (vertical axes) and the simulated RMSEs calculated by the statistical machine learning based surrogate model (horizontal axes) in (a) the Site I, (b) the Site II, and (c) the Site III. See section 5.1.

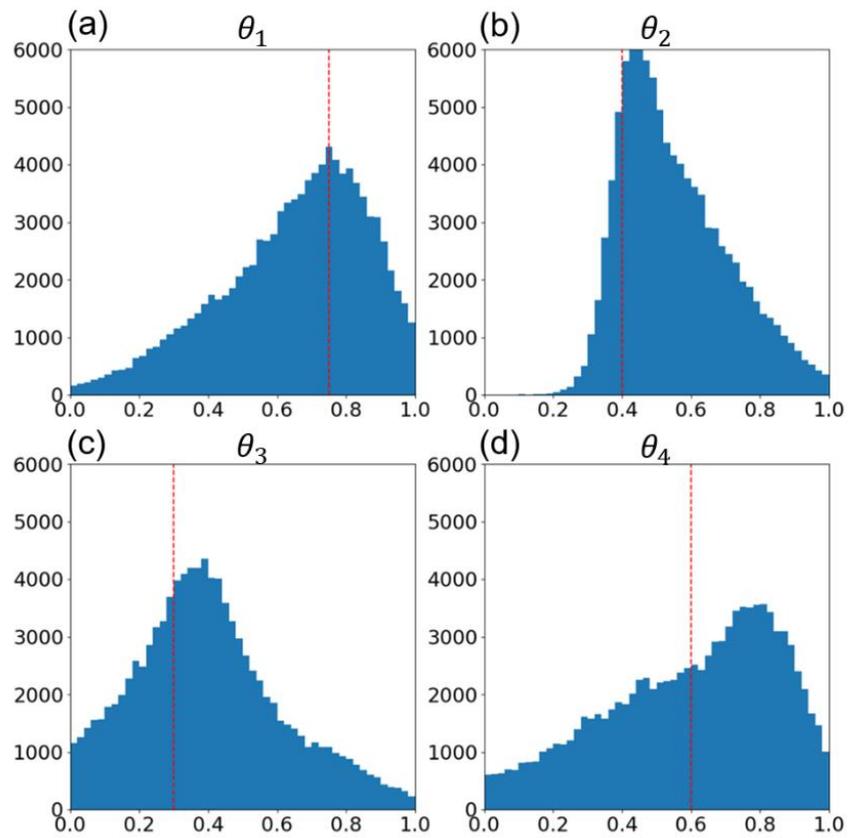
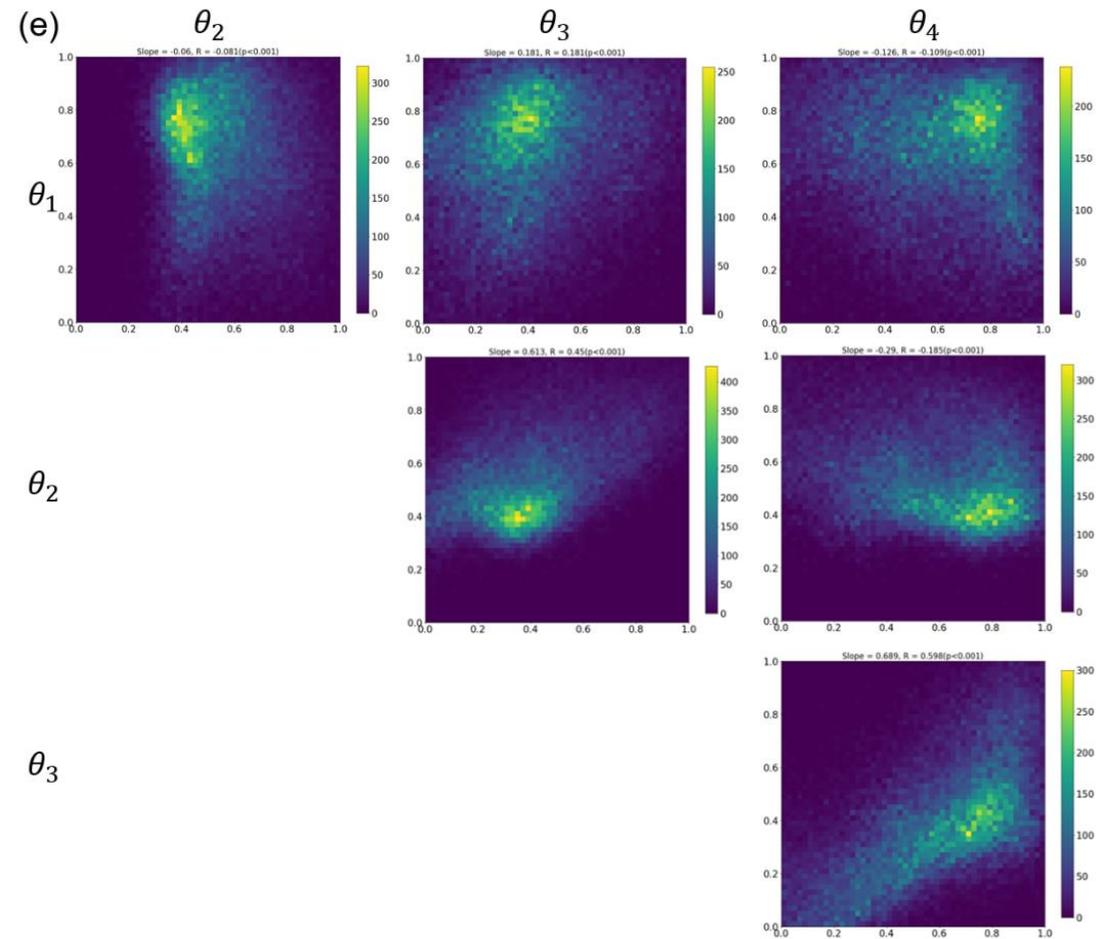

**Figure 4.** The MCMC-sampled posterior probabilistic distributions of the normalized parameters of (a) saturate hydraulic conductivity, (b) the van Genuchten's n, (c) maximum Rubisco efficiency at the leaf top, and (d) the factor controlling the relation between the carbon pools in the Site I. (e) The joint distribution of the each combination of the two parameters in the Site I.

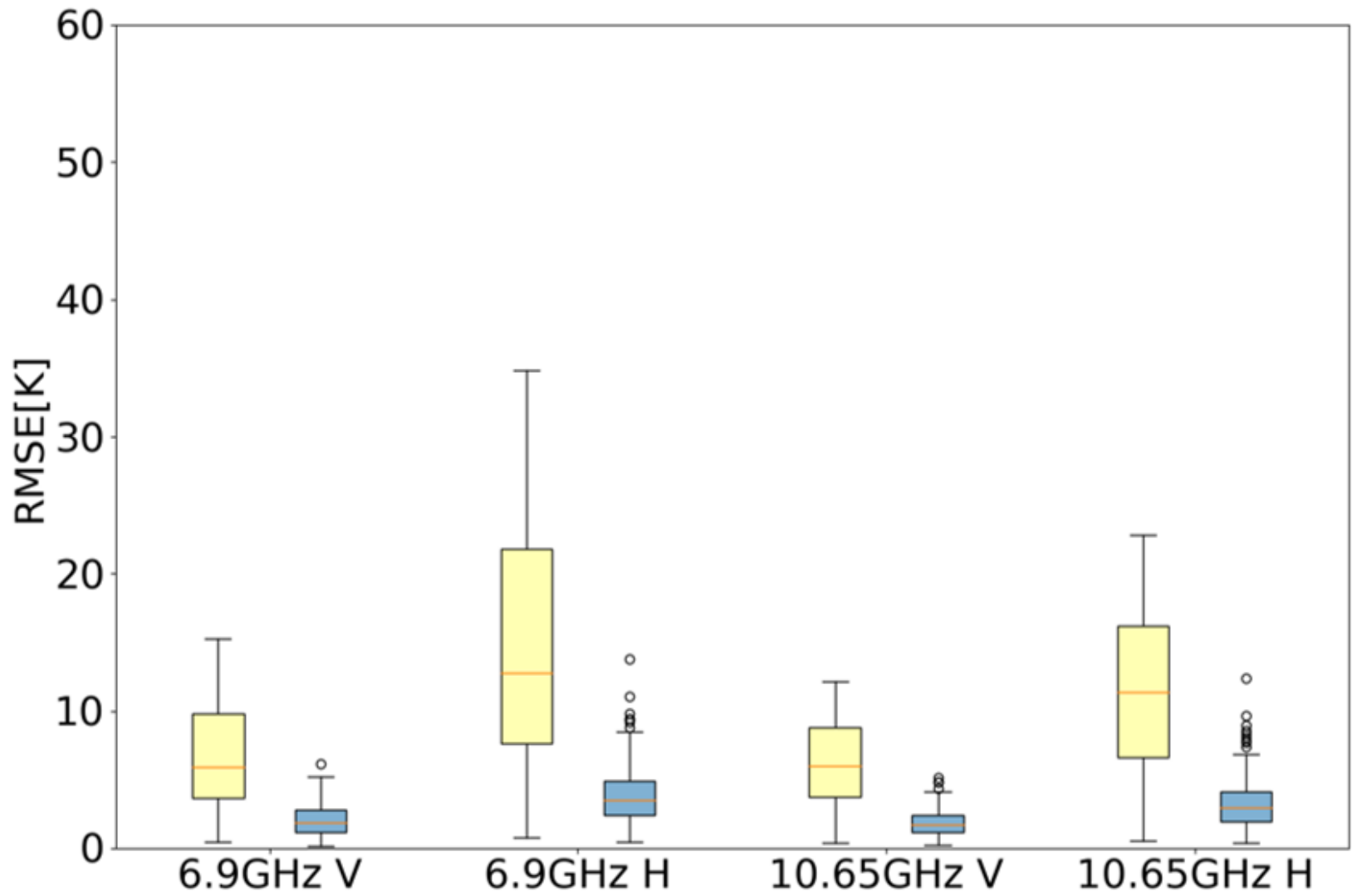

Figure 5. Boxplots of the RMSE of horizontally and vertically polarized brightness temperatures at 6.9GHz and 10.65GHz between the synthetic true observations and the simulated observations by 400 ensembles with the prior parameters (yellow) and the MCMC-sampled parameters (blue) in the Site I.

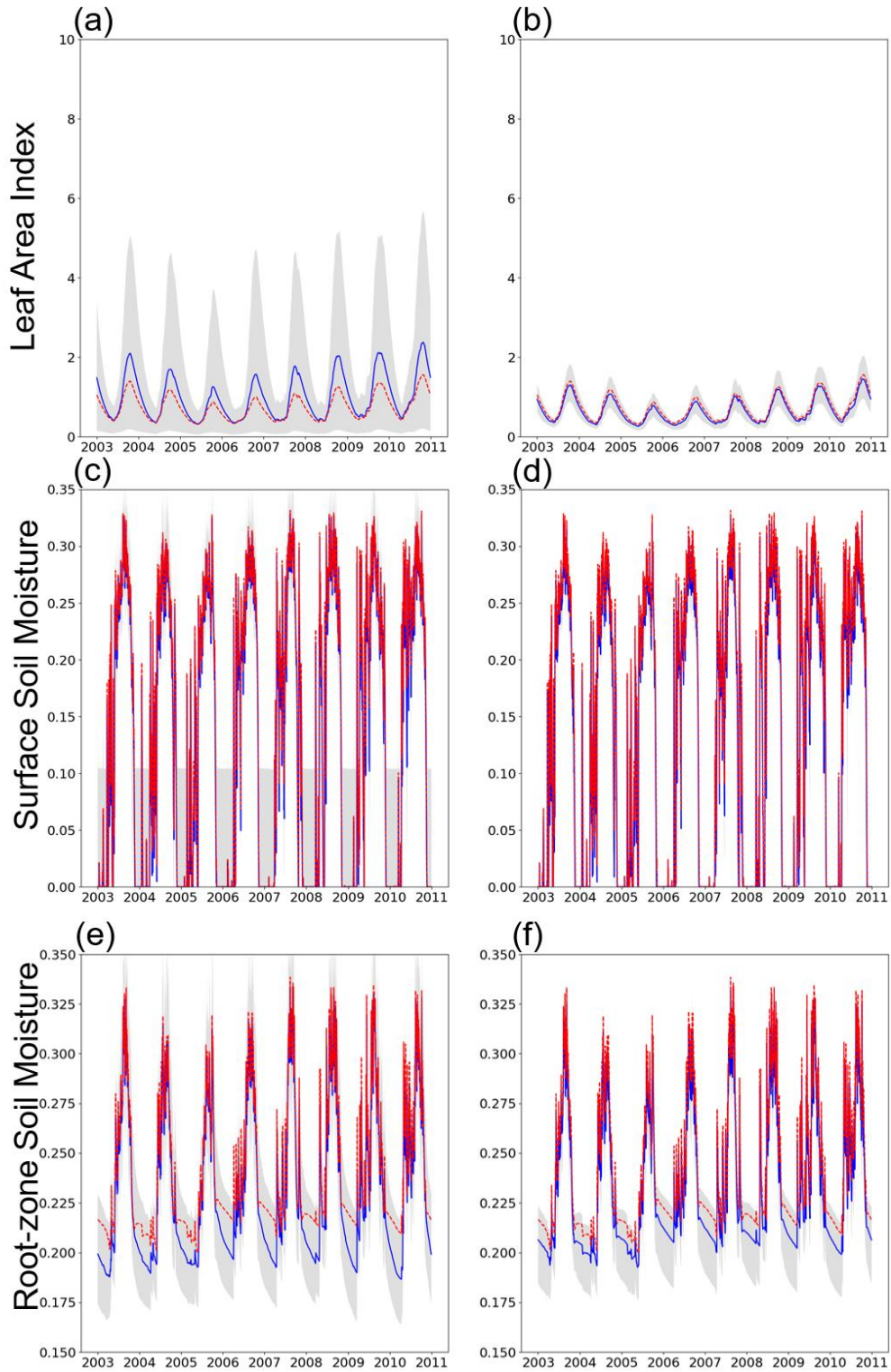

**Figure 6.** Timeseries of (a-b) LAI, (c-d) surface soil moisture, and (e-f) root-zone (35-45cm depth) soil moisture. Red dashed lines are the synthetic truth. Blue lines are the median of 400 ensembles, and grey areas show 5-95% ranges simulated by the prior parameters (a, c, e) and the MCMC-sampled posterior parameters (b, d, f)

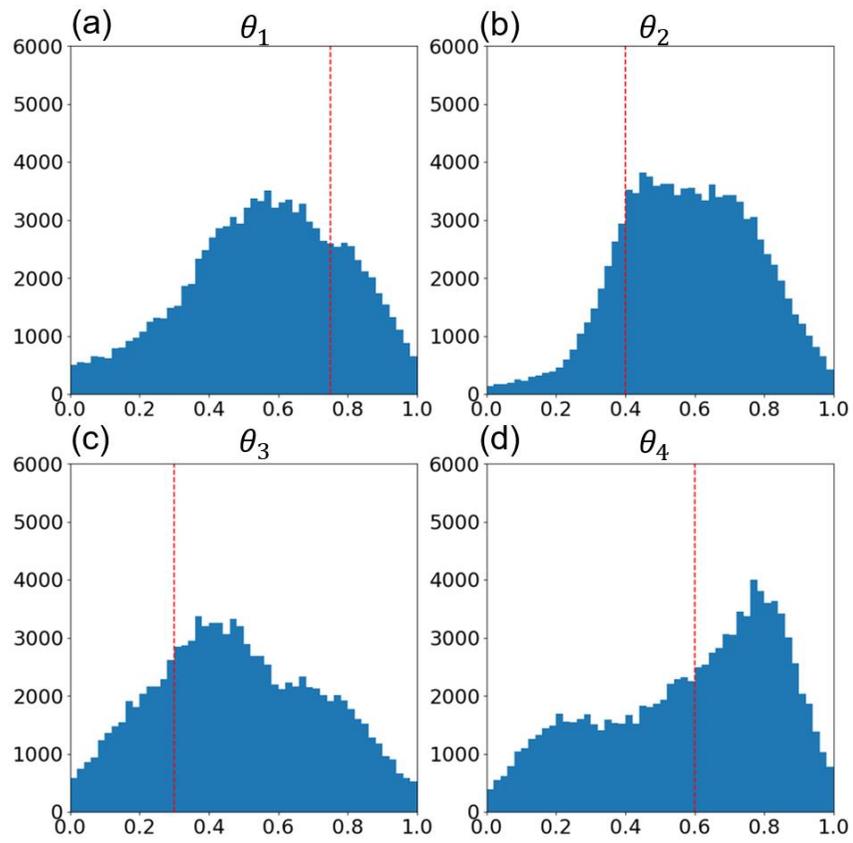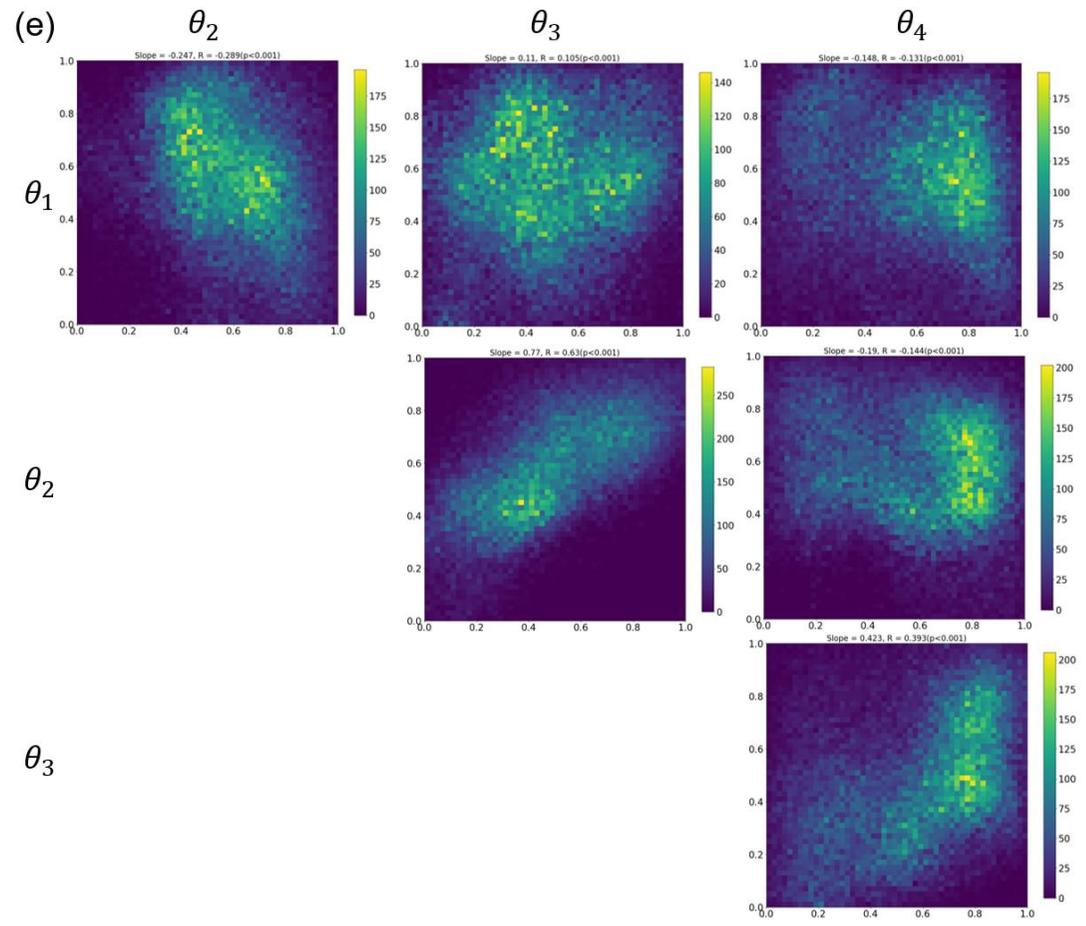

**Figure 7.** Same as Figure 4 but for the Site II.

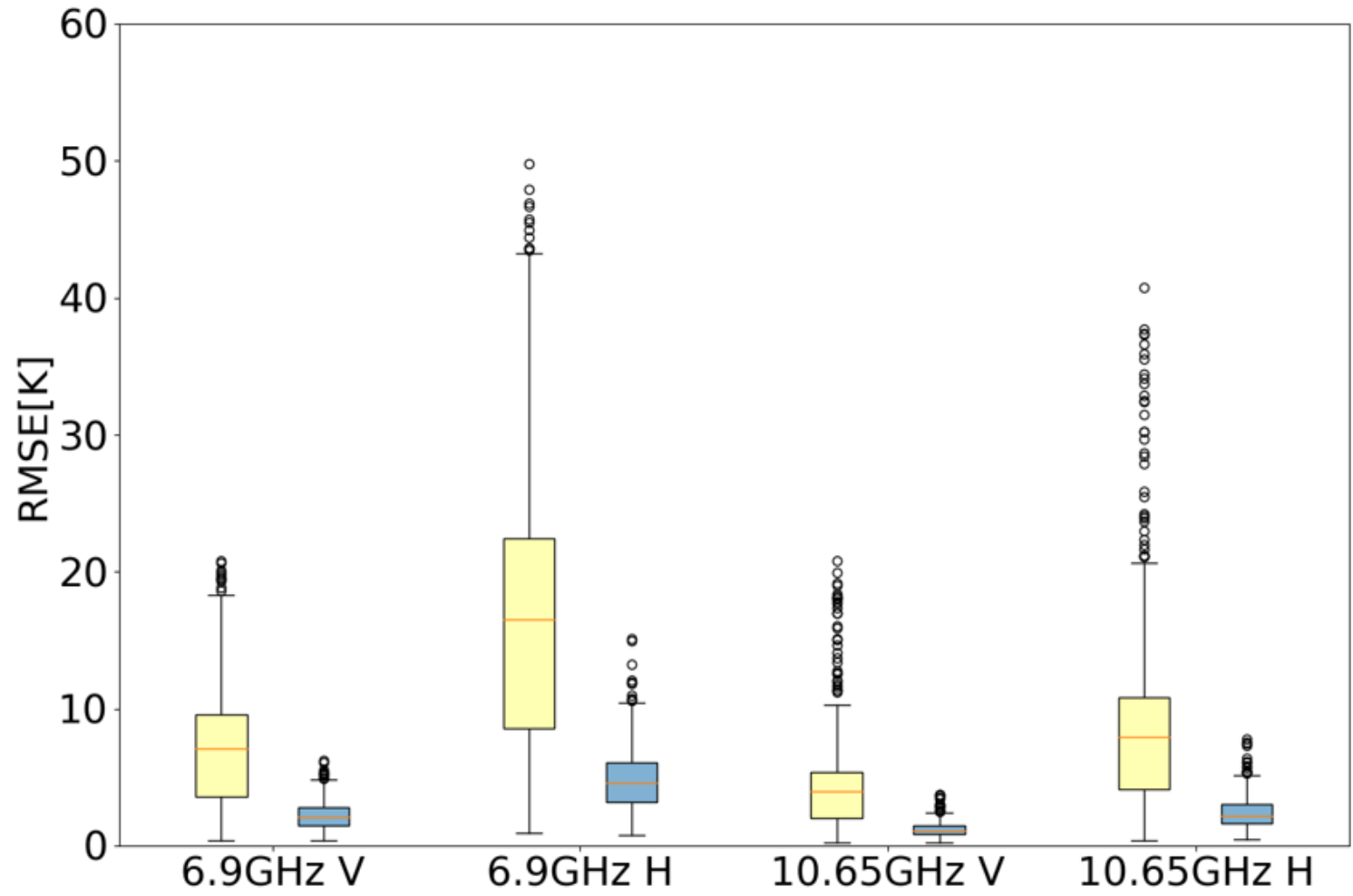

Figure 8. Same as Figure 5 but for the Site II.

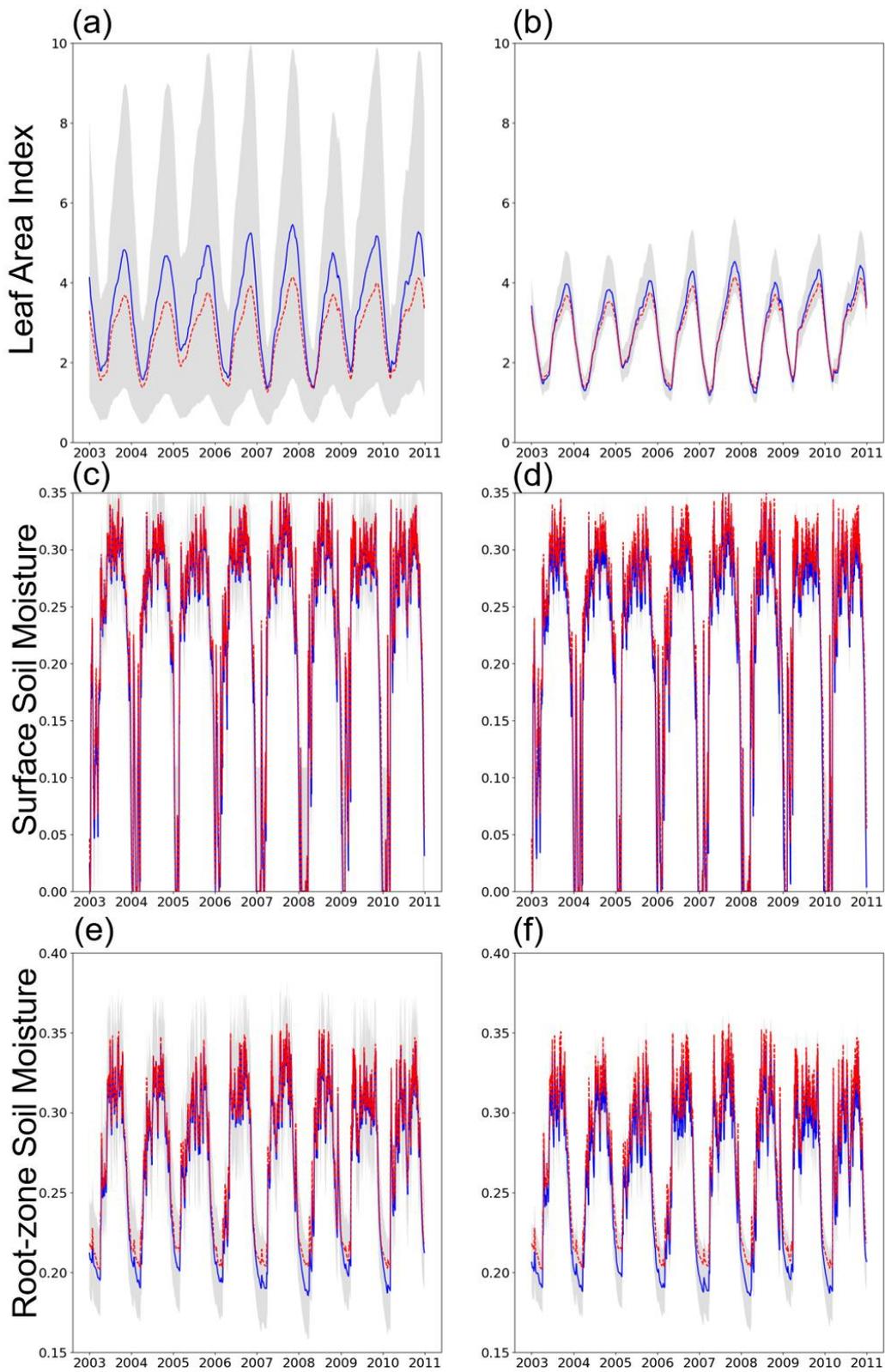

**Figure 9.** Same as Figure 6 but for the Site II

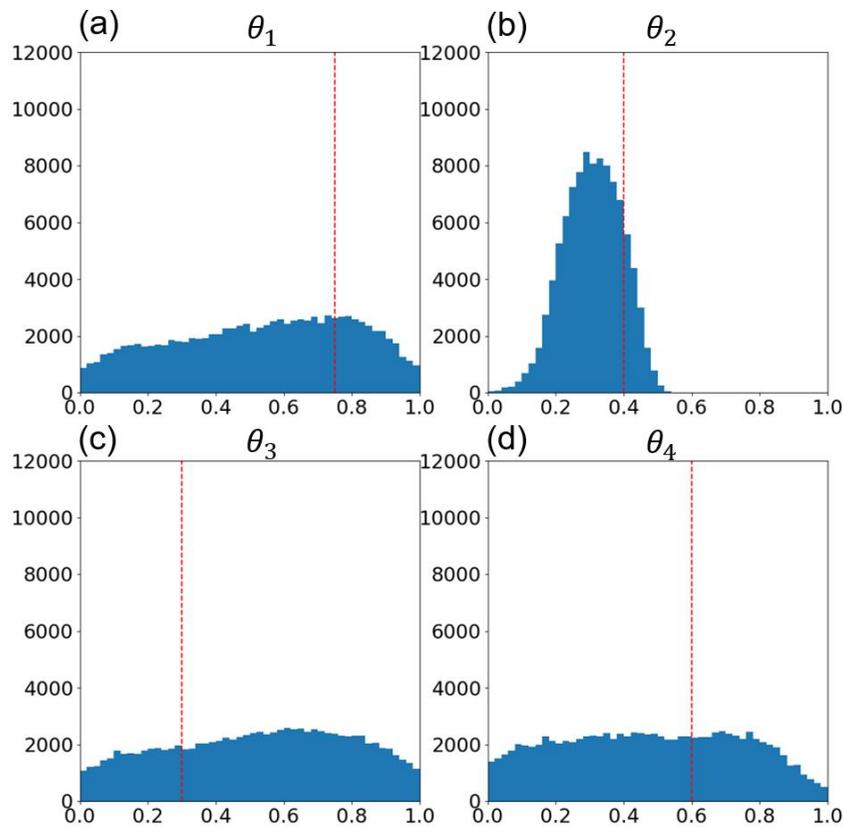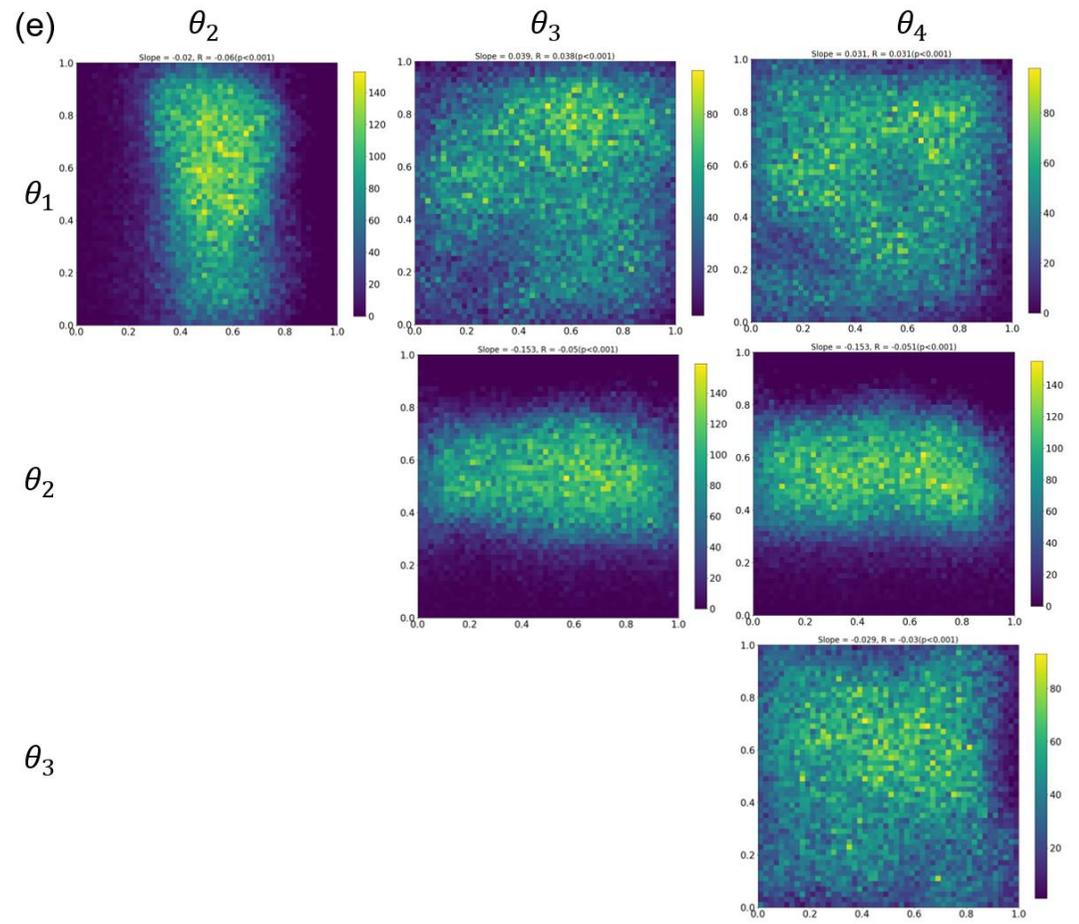

Figure 10. Same as Figure 4 but for the Site III.

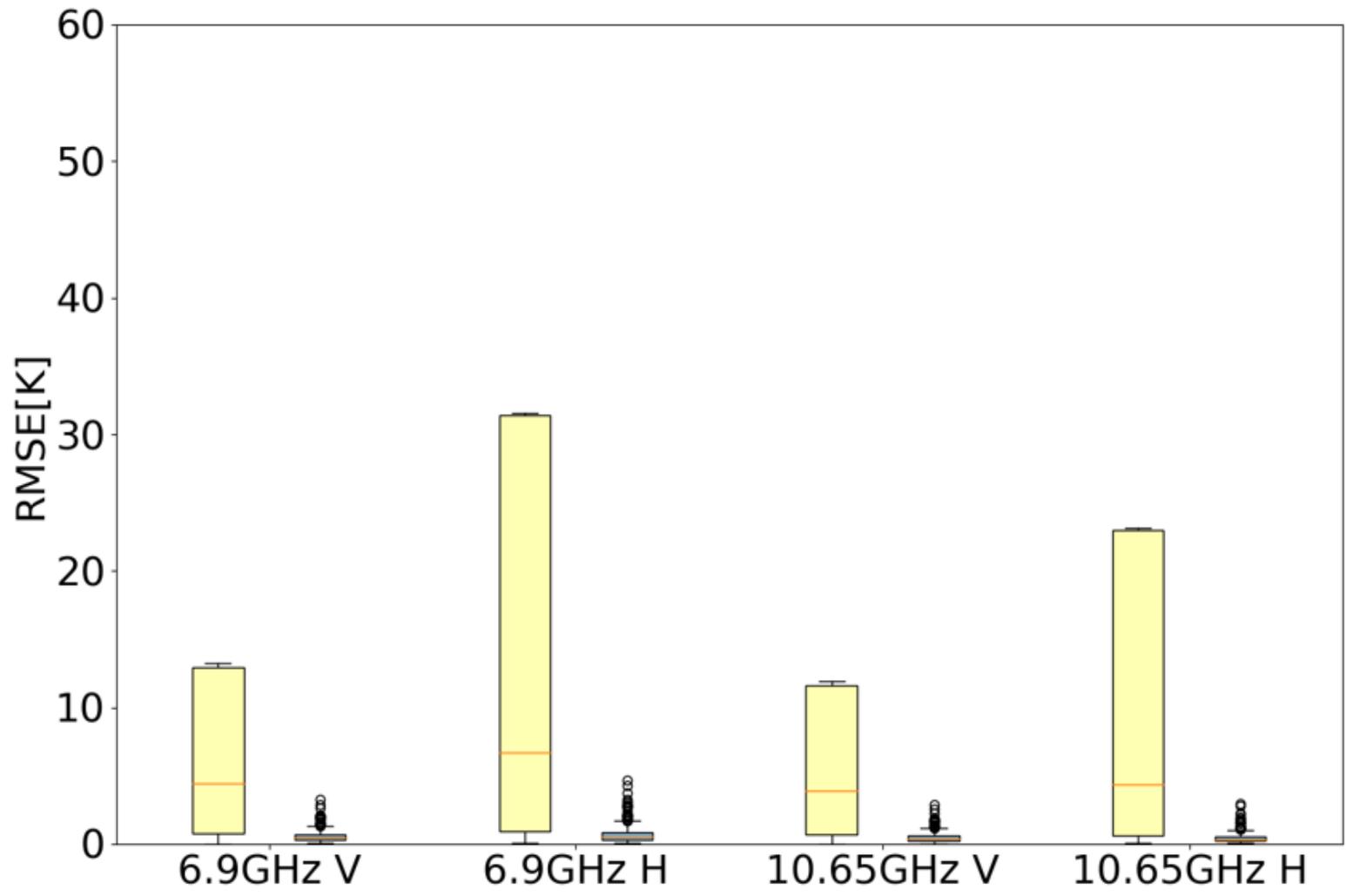

**Figure 11.** Same as Figure 5 but for the Site III.

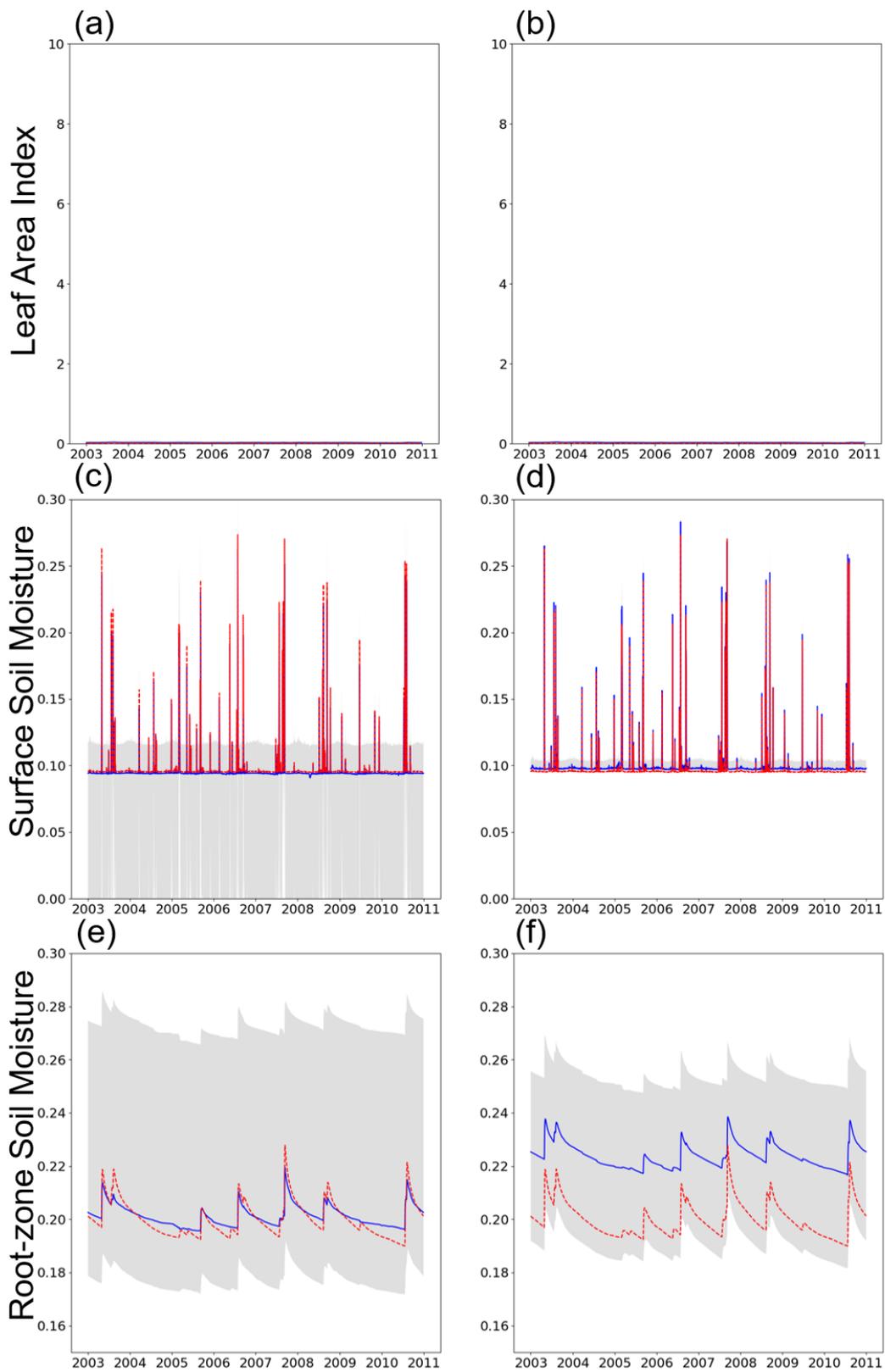

**Figure 12**. Same as Figure 6 but for the Site III.

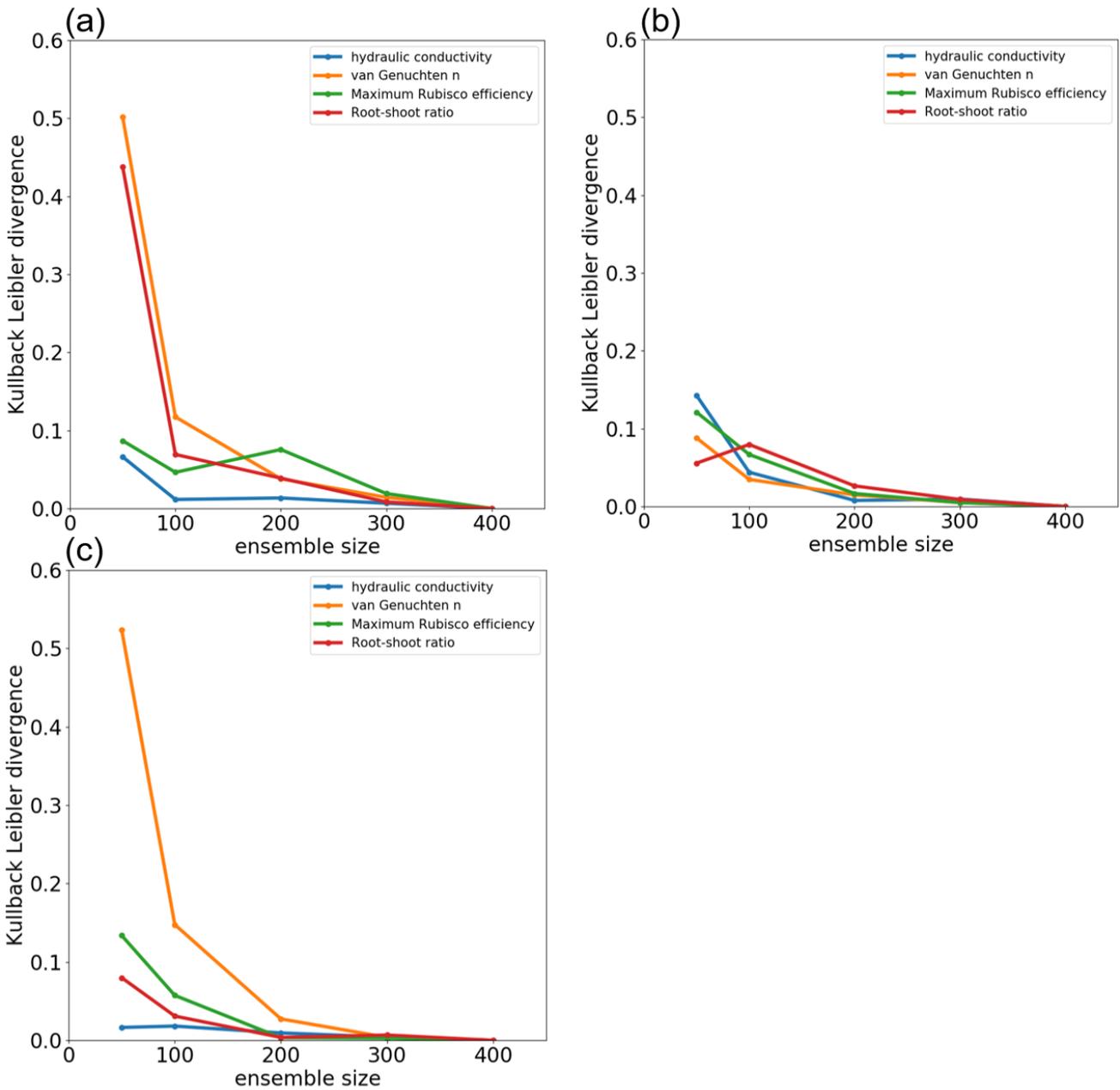

**Figure 13**. Kullback Leibler divergence between the posterior distributions sampled by MCMC with the surrogate model generated by 50, 100, 200, and 300 ensemble members and those by 400 ensemble members for the normalized parameters of saturated hydraulic conductivity (blue), the van Genuchten's n (orange), maximum Rubisco efficiency at the leaf top (green), and the factor controlling the relation between the carbon pools (red) in (a) the Site I, (b) the Site II, and (c) the Site III.

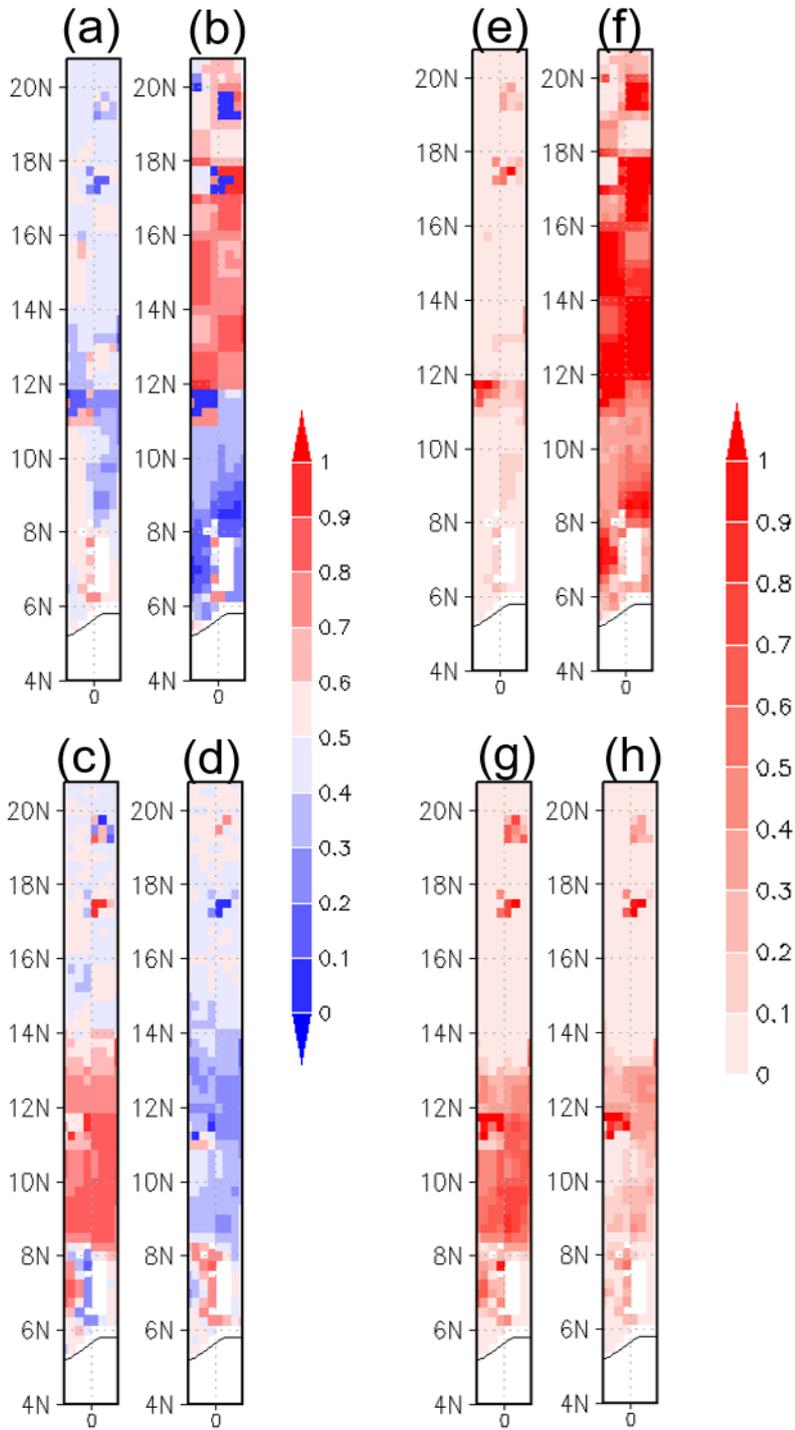

Figure 14. (a-d) Medians of the MCMC-sampled posterior distributions of the normalized parameters of (a) saturate hydraulic conductivity, (b) the van Genuchten's n, (c) maximum Rubisco efficiency at the leaf top, and (d) the factor controlling the relation between the carbon pools. (e-h) The Kullback-Leiblur divergences between the MCMC-sampled distributions and the uniform distributions (see also section 5.2) of (e) saturate hydraulic conductivity, (f) the van Genuchten's n, (g) maximum Rubisco efficiency at the leaf top, and (h) the factor controlling the relation between the carbon pools.

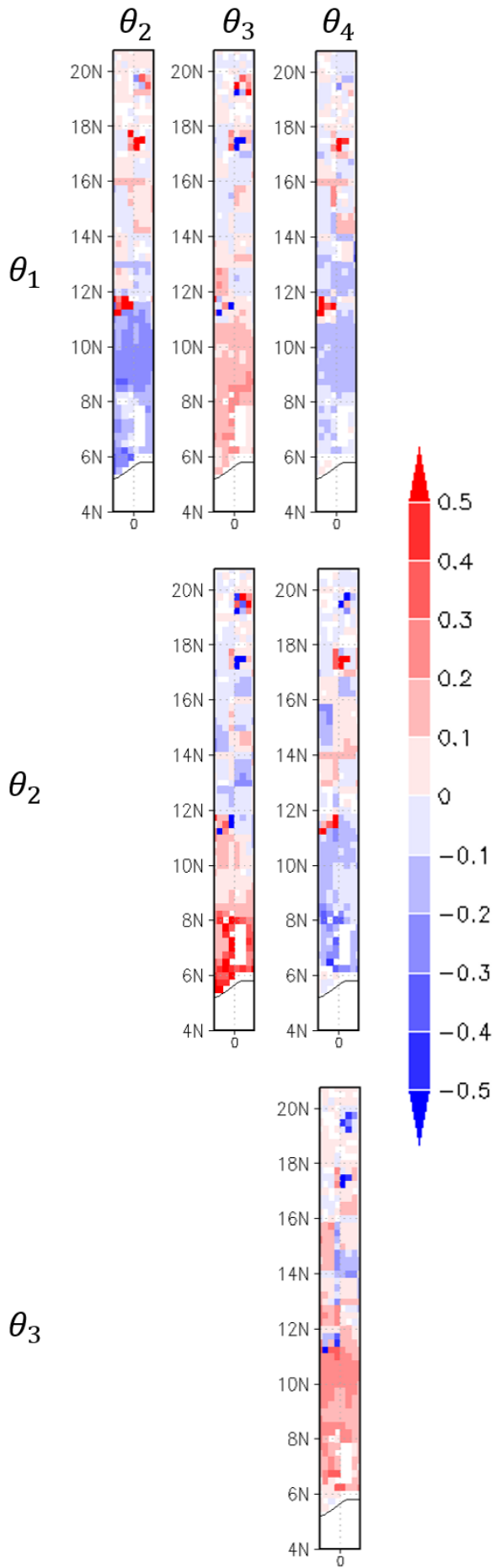

**Figure 15.** The correlation coefficients between the MCMC-sampled distributions of each combination of two parameters.

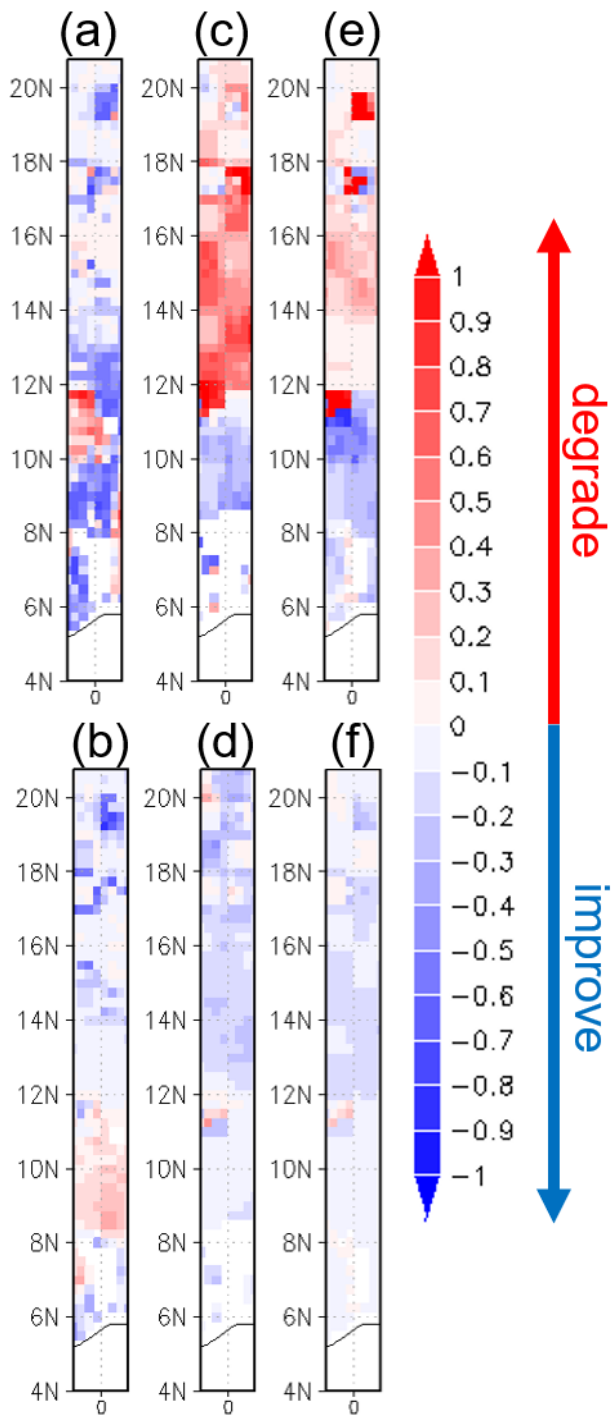

**Figure 16**. The improvement rate (see section 5.2) of (a, c, e) bias scores and (b, d, f) unbiased RMSE scores for (a, b) LAI, (c, d) surface soil moisture, and (e, f) evapotranspiration.

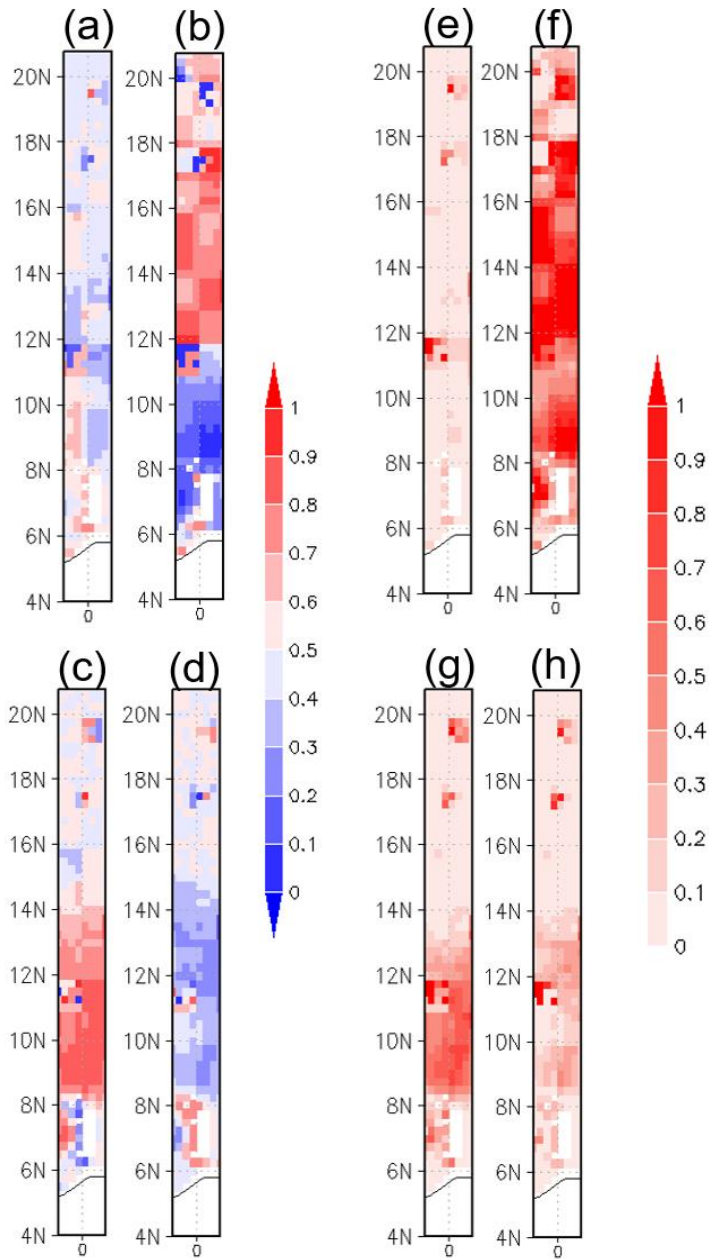

**Figure S1.** Same as Figure 14 but for the experiment in which the study period was split into the 3-year training period and the 5-year validation period. (a-d) Medians of the MCMC-sampled posterior distributions of the normalized parameters of (a) saturate hydraulic conductivity, (b) the van Genuchten's n, (c) maximum Rubisco efficiency at the leaf top, and (d) the factor controlling the relation between the carbon pools. (e-h) The Kullback-Leiblur divergences between the MCMC-sampled distributions and the uniform distributions (see also section 5.2) of (e) saturate hydraulic conductivity, (f) the van Genuchten's n, (g) maximum Rubisco efficiency at the leaf top, and (h) the factor controlling the relation between the carbon pools.

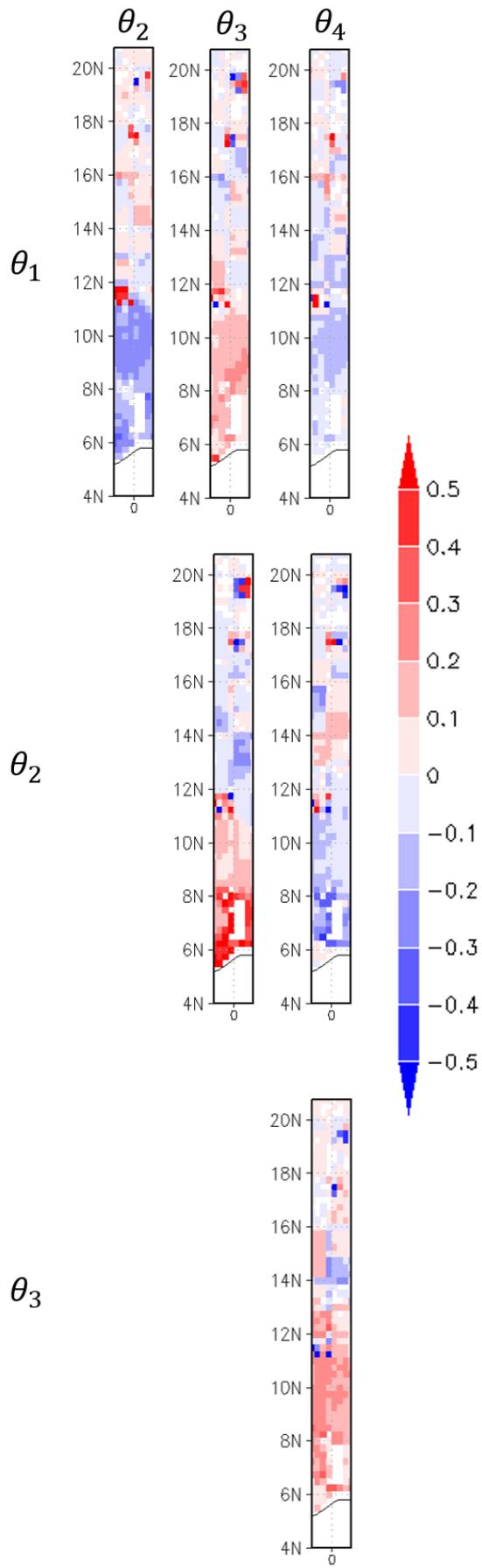

**Figure S2.** Same as Figure 15 but for the experiment in which the study period was split into the 3-year training period and the 5-year validation period. The correlation coefficients between the MCMC-sampled distributions of each combination of two parameters.

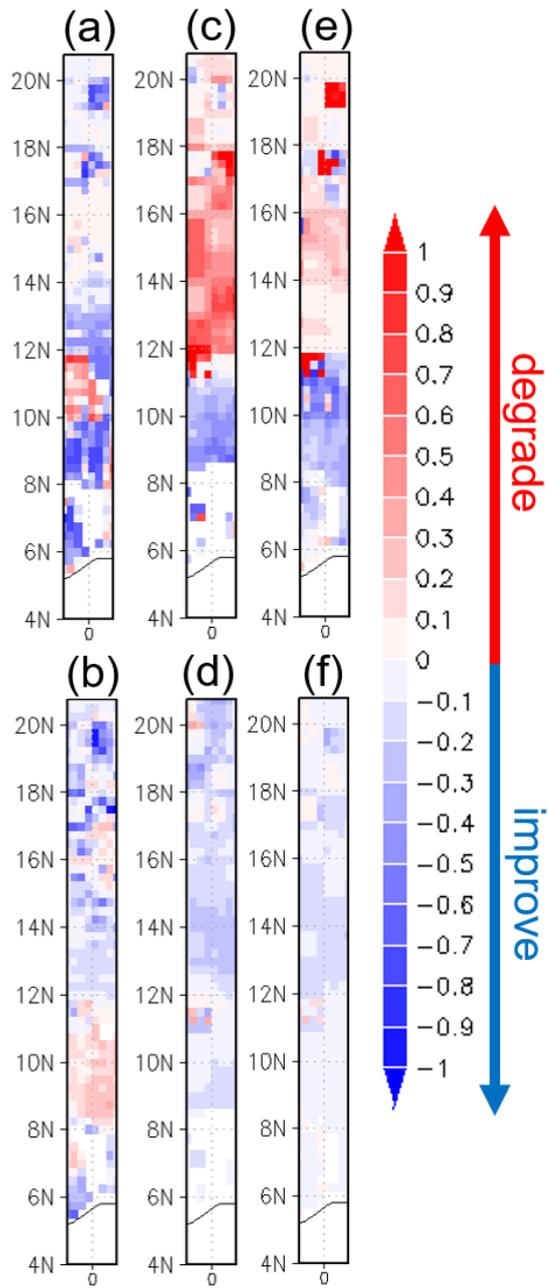

**Figure S3**. Same as Figure 16 but for the experiment in which the study period was split into the 3-year training period and the 5-year validation period. The improvement rate (see section 5.2) of (a, c, e) bias scores and (b, d, f) unbiased RMSE scores for (a, b) LAI, (c, d) surface soil moisture, and (e, f) evapotranspiration.